# A Novel Class of Phase Space Representations for the Exact Population Dynamics of Two-State Quantum Systems and the Relation to Triangle Window Functions

Xiangsong Cheng[1], Xin He[1], Jian Liu[1, *]

[1]*Beijing National Laboratory for Molecular Sciences, Institute of Theoretical and Computational Chemistry, College of Chemistry and Molecular Engineering, Peking University, Beijing 100871, China*

*To whom correspondence should be addressed: jianliupku@pku.edu.cn

**Abstract**

Isomorphism of the two-state system is heuristic in understanding the dynamical or statistical behavior of the simplest yet most quantum system that has no classical counterpart. We use the constraint phase space developed in *J. Chem. Phys.* **2016**, 145, 204105; **2019**, 151, 024105 and *J. Phys. Chem. Lett.* **2021**, 12, 2496-2501, non-covariant phase space functions, time-dependent weight functions, and time-dependent normalization factors to construct a novel class of phase space representations of the exact population dynamics of the two-state quantum system. The equations of motion of the trajectory on constraint phase space are isomorphic to the time-dependent Schrödinger equation. The contribution of each trajectory to the integral expression for the population dynamics is always positive semi-definite. We also prove that the triangle window function approach, albeit proposed as a heuristic empirical model in *J. Chem. Phys.* **2016**, 145, 144108, is related to a special case of the novel class and leads to an isomorphic representation of the exact population dynamics of the two-state quantum system.

I. **Introduction**

Similar to the harmonic oscillator, the two-state system (TSS) has been treated by theoretical physicists and chemists "in ever-increasing levels of abstraction".[1] TSS offers a fundamental paradigmatic model for gaining insight in and teaching quantum mechanics. Some pedagogical examples of TSS include a spin-1/2 particle in a magnetic field, the flip-flop of the nitrogen atom of the $NH_3$ molecule, the polarization states of the photon, the neutral K-meson, the electron transfer between two sites in solution and in biological environment, and so forth[2-6]. Simple as it is, TSS encodes important subtleties of quantum mechanics in many regards. The statistical behavior of TSS is isomorphic to that of the classical one-dimensional Ising model[3]. The dynamical behavior of TSS is isomorphic to that of the Bloch sphere for the spin 1/2 case[7] or the Poincaré sphere for the polarization states of the photon[8], which is related to the $SU(2)$ Stratonovich phase space representation[9]. More recently, the exact constraint coordinate-momentum phase space (CPS) formulation of the finite-state quantum system[10-18] and our forthcoming paper indicate that dynamics of TSS is isomorphic to the linear equations of motion (EOMs) on the quotient space $U(F)/U(F-1)$ with mapping coordinate-momentum variables[12, 16], a special case of the complex Stiefel manifold[19, 20] for $F=2$. Diversified isomorphism structures of TSS are indispensable in understanding the simplest but most quantum model with no true classical analog.

Interestingly, even for the multi-state quantum system, Dirac demonstrated in 1927 that the time-dependent Schrödinger equation (TDSE) of such a system is isomorphic to Hamilton's EOMs for the action-angle variables[21]. In 1979 Meyer and Miller proposed a heuristic mapping Hamiltonian model with the "Langer correction" for the multi-electronic-state molecular system, in which both nuclear and electronic degrees of freedom (DOFs) are treated on the same footing for real-time dynamics[22]. This celebrated Meyer-Miller mapping Hamiltonian model maps a coupled $F$-electronic-state Hamiltonian operator

$$\hat{H} = \sum_{n,m=1}^{F} H_{nm}(\mathbf{R},\mathbf{P})|n\rangle\langle m| \qquad (1)$$

onto a $2F$-dimensional Cartesian phase space $\{\mathbf{x},\mathbf{p}\}=\{x^{(1)},\cdots,x^{(F)},p^{(1)},\cdots,p^{(F)}\}$, that is,

$$H_{map}(\mathbf{x},\mathbf{p};\mathbf{R},\mathbf{P}) = \sum_{n,m=1}^{F}\left[\frac{1}{2}\left(x^{(n)}x^{(m)}+p^{(n)}p^{(m)}\right)-\gamma\delta_{nm}\right]H_{nm}(\mathbf{R},\mathbf{P}) \quad, \qquad (2)$$

where $\{\mathbf{R},\mathbf{P}\}$ are the coordinate and momentum variables for the nuclear DOFs, and $\gamma$ is the parameter accounting for the "zero point energy" (ZPE) for each mapping electronic DOF. (Parameter $\gamma=1/2$ in the original work of Meyer and Miller[22].) The convention $\hbar=1$ is used for discrete (electronic state) DOFs throughout the present paper. The relation between the real action-angle variables $\{\mathbf{N},\boldsymbol{\theta}\}=\{N^{(1)},\cdots,N^{(F)},\theta^{(1)},\cdots,\theta^{(F)}\}$ and the real coordinate-momentum variables $\{\mathbf{x},\mathbf{p}\}$ is

$$\begin{aligned}x^{(j)}+ip^{(j)} &= \sqrt{2(N^{(j)}+\gamma)}\exp(i\theta^{(j)}) \\ &= \sqrt{2e^{(j)}}\exp(i\theta^{(j)}) \qquad j\in\{1,\cdots,F\}\end{aligned} \qquad (3)$$

For convenience, $\left\{e^{(j)} = N^{(j)}+\gamma = \frac{1}{2}\left(\left(x^{(j)}\right)^2+\left(p^{(j)}\right)^2\right)\right\}$ can be used as another set of action variables[23]. In Eq. (3) each angle variable $\theta^{(j)} \in [0, 2\pi)$ for $j \in \{1, \cdots, F\}$. The work of Dirac[21] is a special case of Eq. (3) with parameter $\gamma = 0$. In 1997 Stock and Thoss employed the Schwinger oscillator theory of angular momentum[24, 25] to prove that the Meyer-Miller mapping Hamiltonian (Eq. (2)) with $\gamma = 1/2$ is an exact model in quantum mechanics[26]. Parameter $\gamma$ is, however, often set to $1/3$, $(\sqrt{3}-1)/2$, and other non-negative values in its semiclassical/quasiclassical applications[26-35]. (As comparison, parameter $\gamma$ is not interpreted as a ZPE parameter in the generalized exact coordinate-momentum formulation of quantum mechanics[10, 12-16] for the finite-state system.) When the linearized semi-classical initial value representation (LSC-IVR) is applied to both nuclear and electronic DOFs to study nonadiabatic processes[27], it effectively employs the infinite Wigner coordinate-momentum phase space representation of both nuclear and electronic DOFs[12, 13, 16]. Such an approach for nonadiabatic systems meets its difficulties. E.g., it leads to relatively poor results for the population-population correlation function (i.e., the correlation function of the population dynamics), which can occasionally become even negative, for the three-state photo-dissociation models[36]; and it fails to capture the bifurcation characteristic of nuclear dynamics in the asymptotic region[37]. Since 2013 Cotton and Miller have introduced window functions to deal with electronic DOFs while describing nuclear DOFs by the Wigner function for the Meyer-Miller Hamiltonian, leading to the symmetrical quasi-classical (SQC) approach that has been applied in exploring various nonadiabatic systems[23, 28-30, 35, 38-50]. The population-population correlation function of the SQC approach for nonadiabatic dynamics is guaranteed to be positive semi-definite when the expression of the initial total density operator of both nuclear and electronic DOFs is non-negative. It is expected that the (reduced) electronic population-population correlation function is positive semi-definite for most of the other realistic nonadiabatic transition cases. The most useful window function is of the triangle form proposed in ref [28], which performs well in both the weak state-coupling region and normal region. It is generally believed that the triangle window functions (TWFs) for describing electronic DOFs in the SQC approach offer an effective but *empirical* model. We focus on only the pure *F*-state quantum system, i.e., the frozen-nuclei limit of Eq. (1), where each element $H_{nm}(\mathbf{R},\mathbf{P}) \equiv H_{nm}$ is constant. That is, the Hamiltonian operator of Eq. (1) becomes

$$\hat{H} = \sum_{n,m=1}^{F} H_{nm} |n\rangle\langle m|.$$

The purpose of the paper is to propose a novel class of isomorphic representations for the exact population dynamics of TSS. It employs time-dependent normalization factors, time-dependent weight functions, and non-covariant phase space functions on the constraint coordinate-momentum phase space formulation of the finite-state quantum system[10, 12-18]. All these components associated with each trajectory on phase space are positive semi-definite. We prove that the triangle window function approach of Ref. [28] is related to a special case of the new isomorphism class and offers an isomorphism for reproducing the exact population dynamics of TSS.

The paper is organized as follows. Section II introduces the phase space mapping formalism on CPS, and derives several useful formulas for the evolution of mapping variables when the EOMs of $\mathbf{x}+i\mathbf{p}$ are isomorphic to the TDSE. Section III presents a generalized phase space formulation. By employing a few variable-transformations of integrals, we show that the exact population-population correlation function can be related to an axial symmetry of the time-dependent normalization factor. The axial symmetry leads to an integral equation as a variant of *Abel equation*[51]. The solution to the integral equation then yields a new class of exact representations of the population-population correlation function (of TSS). In the integral expression of the population-population correlation function, the value contributed by each trajectory is always positive semi-definite by construction. Section IV focuses on a special case, where the original TWF expression of electronic DOFs of the SQC approach is transformed to an equivalent integral form on CPS of TSS. By showing that the form falls into the new class of phase space representations of Section III, we prove that the SQC-TWF approach produces the exact population-population correlation function of TSS. Finally, conclusion remarks are presented in Section V.

## II. Theoretical Background

### II-A. Constraint Coordinate-Momentum Phase Space Formulation of the Finite-State Quantum System

The *constraint* coordinate-momentum *phase space* (CPS) formulation, which has recently been developed based on Refs. [10] and [12], offers a one-to-one correspondence mapping between any quantum operator $\hat{A}$ of the $F$-state system and its corresponding phase space function,

$$A_C(\mathbf{x},\mathbf{p};\mathbf{\Gamma}) = \text{Tr}_e\left[\hat{A}\hat{K}_{\text{ele}}(\mathbf{x},\mathbf{p};\mathbf{\Gamma})\right] , \qquad (4)$$

with coordinate-momentum variables $(\mathbf{x},\mathbf{p})$ and commutator matrix variables[14] $\{\Gamma_{mn}\}$. Here, $\mathrm{Tr}_e[\ ]$ represents the trace over the $F$ states. The mapping kernel[13-15, 52] $\hat{K}_{\mathrm{ele}}(\mathbf{x},\mathbf{p};\Gamma)$ on CPS reads

$$\hat{K}_{\mathrm{ele}}(\mathbf{x},\mathbf{p},\Gamma) = \sum_{n,m=1}^{F}\left[\frac{1}{2}\left(x^{(n)}+ip^{(n)}\right)\left(x^{(m)}-ip^{(m)}\right)-\Gamma_{nm}\right]|n\rangle\langle m|, \quad (5)$$

where the mathematical structure of CPS is related to the complex Stiefel manifold[19, 20], i.e., the quotient space $\mathrm{U}(F)/\mathrm{U}(F-r)$ with $1 \leq r < F$. In the simplest case $(r=1)$, commutator matrix $\Gamma = \gamma \mathbf{1}$ is a constant matrix, and the mapping CPS is the $\mathrm{U}(F)/\mathrm{U}(F-1)$ constraint (coordinate-momentum) phase space[10, 12, 13, 16],

$$S(\mathbf{x},\mathbf{p};\gamma) = \delta\left(\sum_{n=1}^{F}\frac{1}{2}\left(\left(x^{(n)}\right)^2+\left(p^{(n)}\right)^2\right)-(1+F\gamma)\right) \quad (6)$$

with parameter $\gamma \in (-1/F, \infty)$. The constraint phase space, Eq. (6), can also be interpreted with action-angle variables as

$$S(\mathbf{e},\boldsymbol{\theta};\gamma) = \delta\left(\sum_{n=1}^{F}e^{(n)}-(1+F\gamma)\right), \quad (7)$$

which was used in Appendix A of Ref. [12] for the $\gamma = 0$ case for general $F$-state systems. Figure 1 shows both the simplex representation [Eq. (7)] and sphere representation [Eq. (6)] of CPS for the $F$-state quantum system, where only the mapping variables corresponding to two states, $i$ and $j$, are demonstrated.

(a) With action-angle variables

(b) With coordinate-momentum variables

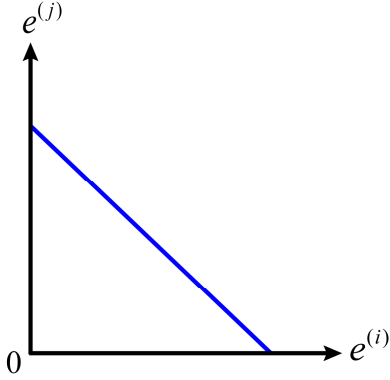
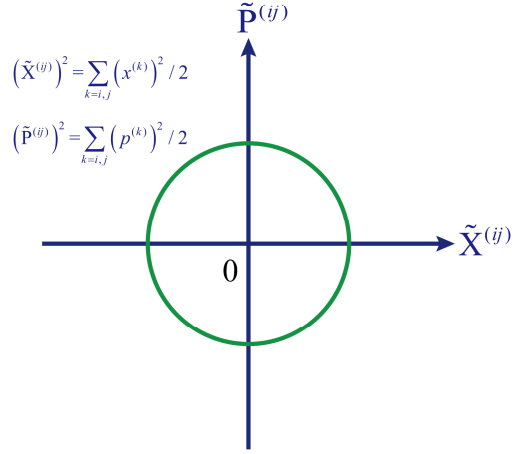

$$e^{(i)} + e^{(j)} = 1 + F\gamma - \sum_{k \neq i,j}^{F} e^{(k)}$$

$$\left(\tilde{X}^{(ij)}\right)^2 + \left(\tilde{P}^{(ij)}\right)^2 = 1 + F\gamma - \sum_{k \neq i,j}^{F} \left[\left(x^{(k)}\right)^2 + \left(p^{(k)}\right)^2\right]/2$$

Simplex Representation of CPS with $e^{(i)}$ and $e^{(j)}$

Sphere Representation of CPS with $x^{(i)}, p^{(i)}, x^{(j)}, p^{(j)}$

**Figure 1.** The simplex and sphere representations of CPS of the mapping variables for states $i$ and $j$ of the multi-state quantum system. Panel (a): The simplex representation of CPS with action-angle variables. The action-angle constraint $\mathcal{S}(\mathbf{e}, \boldsymbol{\theta}; \gamma)$ for general multi-state systems was firstly proposed in Eqs. (A4)-(A5) of Appendix A of Ref. [12] where $\gamma = 0$. The variables used in the figure are $e^{(i)}$ and $e^{(j)}$. Panel (b): The sphere representation of CPS with coordinate-momentum variables. Constraint $\mathcal{S}(\mathbf{x}, \mathbf{p}; \gamma)$ for general multi-state systems was first proposed in Eqs. (5) and (28) of Ref. [12] and later in Eq. (4) of Ref. [13]. The variables employed in the figure are $\tilde{X}^{(ij)}$ and $\tilde{P}^{(ij)}$ derived from $x^{(i)}$, $p^{(i)}$, $x^{(j)}$ and $p^{(j)}$. Equations (A13), (A14), (A16) and (A20) with action-angle variables of Appendix A of Ref. [12] presented the rigorous expression on CPS for parameter $\gamma = 0$ for population dynamics for general $F$-state systems, which was straightforwardly generalized to Eq. (20) of Ref. [13]

with coordinate-momentum variables on CPS for parameter $\gamma \in (-1/F, \infty)$.

The mapping kernel on the $\mathrm{U}(F)/\mathrm{U}(F-1)$ CPS[10, 12, 13, 16] is

$$\hat{K}_{\mathrm{ele}}(\mathbf{x},\mathbf{p};\gamma) = \sum_{n,m=1}^{F} \left[ \frac{1}{2}\left(x^{(n)} + ip^{(n)}\right)\left(x^{(m)} - ip^{(m)}\right) - \gamma \delta_{nm} \right] |n\rangle\langle m| \quad . \tag{8}$$

As shown in Refs. [12-15], the mapping phase space function of the Hamiltonian operator Eq. (1), rigorously derived from the mapping relation Eq. (4) with the kernel Eq. (8) on the $\mathrm{U}(F)/\mathrm{U}(F-1)$ quotient space defined by Eq. (6), is reminiscent of the well-known Meyer-Miller mapping Hamiltonian[22], Eq. (2). More recently, the correct relation between the $\mathrm{SU}(2)$ or $\mathrm{SU}(F)$ Stratonovich phase space and the $\mathrm{U}(F)/\mathrm{U}(F-1)$ constraint coordinate-momentum phase space[10, 12, 13] has been pointed out in Refs. [14, 15] and explicitly presented in Ref. [16].

The phase space representation of the trace of the product of two operators reads

$$\mathrm{Tr}_e[\hat{A}\hat{B}] = \int_{\mathcal{S}(\mathbf{x},\mathbf{p};\gamma)} \mathrm{d}\boldsymbol{\mu}_{\mathrm{ele}}(\mathbf{x},\mathbf{p};\gamma) A_C(\mathbf{x},\mathbf{p};\gamma) \tilde{B}_C(\mathbf{x},\mathbf{p};\gamma) \tag{9}$$

with

$$\tilde{B}_C(\mathbf{x},\mathbf{p};\gamma) = \mathrm{Tr}_e\left[\hat{K}_{\mathrm{ele}}^{-1}(\mathbf{x},\mathbf{p};\gamma)\hat{B}\right] \quad . \tag{10}$$

Here,

$$\mathrm{d}\boldsymbol{\mu}_{\mathrm{ele}}(\mathbf{x},\mathbf{p};\gamma) = F\mathrm{d}\mathbf{x}\mathrm{d}\mathbf{p} \tag{11}$$

is the integration measure on CPS, and

$$\int_{\mathcal{S}(\mathbf{x},\mathbf{p};\gamma)} F\mathrm{d}\mathbf{x}\mathrm{d}\mathbf{p}\, g(\mathbf{x},\mathbf{p}) = \int F\mathrm{d}\mathbf{x}\mathrm{d}\mathbf{p}\, \frac{1}{\Omega(\gamma)} \mathcal{S}(\mathbf{x},\mathbf{p};\gamma) g(\mathbf{x},\mathbf{p}), \tag{12}$$

where [12, 13, 16]

$$\Omega(\gamma) = \int d\mathbf{x} d\mathbf{p}\, \mathcal{S}(\mathbf{x},\mathbf{p};\gamma) = \frac{(2\pi)^F (1+F\gamma)^{F-1}}{(F-1)!}. \qquad (13)$$

The inverse mapping kernel that satisfies the exact mapping condition

$$\begin{cases} \hat{K}_{ele}(\mathbf{x},\mathbf{p};\gamma) = \sum_{n,m=1}^{F} K_{nm}(\mathbf{x},\mathbf{p};\gamma)|n\rangle\langle m| \\ \hat{K}_{ele}^{-1}(\mathbf{x},\mathbf{p};\gamma) = \sum_{n,m=1}^{F} K_{nm}^{-1}(\mathbf{x},\mathbf{p};\gamma)|n\rangle\langle m| \\ \int_{\mathcal{S}(\mathbf{x},\mathbf{p};\gamma)} F d\mathbf{x} d\mathbf{p}\, K_{mn}(\mathbf{x},\mathbf{p};\gamma) K_{lk}^{-1}(\mathbf{x},\mathbf{p};\gamma) = \delta_{mk}\delta_{nl} \end{cases} \qquad (14)$$

is *not* unique[17]. A special choice of refs [12-14] is

$$\hat{K}_{ele}^{-1}(\mathbf{x},\mathbf{p};\gamma) = \sum_{n,m=1}^{F}\left[\frac{1+F}{2(1+F\gamma)^2}\left(x^{(n)}+ip^{(n)}\right)\left(x^{(m)}-ip^{(m)}\right) - \frac{1-\gamma}{1+F\gamma}\delta_{nm}\right]|n\rangle\langle m|. \qquad (15)$$

When operator $\hat{B}$ is replaced by its time-dependent Heisenberg operator $\hat{B}(t)$, the evaluation of Eq. (9), which becomes the time correlation function, is replaced by the *exact* trajectory-based dynamics on CPS for the finite-state quantum system,

$$\begin{aligned} \text{Tr}_e\left[\hat{A}(0)\hat{B}(t)\right] &= \text{Tr}_e\left[\hat{A}e^{i\hat{H}t}\hat{B}e^{-i\hat{H}t}\right] \\ &= \int_{\mathcal{S}(\mathbf{x},\mathbf{p};\gamma)} d\mu_{ele}(\mathbf{x},\mathbf{p};\gamma) A_C(\mathbf{x},\mathbf{p};\gamma)\tilde{B}_C(\mathbf{x},\mathbf{p};\gamma;t) \\ &= \int_{\mathcal{S}(\mathbf{x}_0,\mathbf{p}_0;\gamma)} d\mu_{ele}(\mathbf{x}_0,\mathbf{p}_0;\gamma) A_C(\mathbf{x}_0,\mathbf{p}_0;\gamma)\tilde{B}_C(\mathbf{x}_t,\mathbf{p}_t;\gamma) \end{aligned}, \qquad (16)$$

where $\hat{K}_{ele}(\mathbf{x},\mathbf{p};\gamma)$ and $\hat{K}_{ele}^{-1}(\mathbf{x},\mathbf{p};\gamma)$ are chosen as Eq. (8) and Eq. (15), and the EOMs of trajectories are isomorphic to the TDSE[12-14]. The EOMs of trajectories will be further discussed in the next subsection. As a special case of Eq. (16), the population-population correlation function on CPS reads

$$\begin{aligned} &\text{Tr}_e\left[|n\rangle\langle n|e^{i\hat{H}t}|m\rangle\langle m|e^{-i\hat{H}t}\right] \\ &= \int_{\mathcal{S}(\mathbf{x}_0,\mathbf{p}_0;\gamma)} d\mu_{ele}(\mathbf{x}_0,\mathbf{p}_0;\gamma) K_{nn}(\mathbf{x}_0,\mathbf{p}_0;\gamma) K_{mm}^{-1}(\mathbf{x}_t,\mathbf{p}_t;\gamma) \end{aligned}. \qquad (17)$$

**II-B. Evolution of Mapping Phase Space Variables on Constraint Phase Space for the Two-State System**

Consider the $F$-state quantum system. Any normalized pure quantum state can be expressed as

$$|\psi(t)\rangle = \sum_{n=1}^{F} \frac{g^{(n)}(t)}{\sqrt{2(1+F\gamma)}} |n\rangle \quad , \tag{18}$$

where the coefficients $\{g^{(n)}\}$ are complex. Denote the coefficient vector $\mathbf{g} \equiv \mathbf{x} + i\mathbf{p}$, where $(\mathbf{x}, \mathbf{p})$ are real (coordinate-momentum) variables. As discussed in Refs. [10, 21, 22], it is straightforward to show that the TDSE of the $F$-state quantum system

$$i\frac{\partial}{\partial t}|\psi(t)\rangle = \hat{H}|\psi(t)\rangle \tag{19}$$

leads to the exact linear EOMs of $(\mathbf{x}_t, \mathbf{p}_t)$

$$\dot{\mathbf{g}}_t = -i\mathbf{H}\mathbf{g}_t \quad , \tag{20}$$

where $\mathbf{H}$ is the $F \times F$ Hamiltonian matrix, of which the element in the $n$-th row and $m$-th column is $H_{nm}$. The solution to Eq. (20) is

$$\mathbf{g}_t = \mathbf{U}(t)\mathbf{g}_0 \equiv e^{-i\mathbf{H}t}\mathbf{g}_0 \quad . \tag{21}$$

When TSS (where $F = 2$) is considered, the time evolution operator $\hat{U}(t) \equiv e^{-i\hat{H}t}$ (of Eq. (16)) is represented by a $2 \times 2$ unitary matrix $\mathbf{U}(t) = e^{-i\mathbf{H}t}$. Matrix $\mathbf{U}(t)$ can be expressed with the elements of the Hamiltonian matrix, $\mathbf{H}$. Following Refs. [53-55], we express $\mathbf{U}(t)$ as

$$\mathbf{U}(t) = e^{-i\Phi} \begin{pmatrix} e^{i\psi}\cos\xi & e^{i\varphi}\sin\xi \\ -e^{-i\varphi}\sin\xi & e^{-i\psi}\cos\xi \end{pmatrix} \quad , \tag{22}$$

where $(\xi, \Phi, \varphi, \psi)$ are functions of $t$ and the elements of $\mathbf{H}$. The explicit expressions of $(\xi, \Phi, \varphi, \psi)$ are presented in Appendix A. When $t = 0$, we set $\Phi = \xi = \psi = 0$. The range of $\xi$ is $[0, \pi/2]$ as discussed around Eq. (96) of Appendix A. The expression of the evolution matrix similar to Eq. (22) with $e^{-i\Phi}$ is often used in the field of quantum computing and information[54, 55].

The initial condition can be represented by either $(\mathbf{x}_0, \mathbf{p}_0)$ or $(\mathbf{e}_0, \boldsymbol{\theta}_0)$. Using the action-angle variables, the 2-dimensional vector at time $t = 0$ of Eq. (21), $\mathbf{g}_0$, reads

$$\mathbf{g}_0 = \begin{pmatrix} \sqrt{2e_0^{(1)}} e^{i\theta_0^{(1)}} \\ \sqrt{2e_0^{(2)}} e^{i\theta_0^{(2)}} \end{pmatrix} . \tag{23}$$

On the constraint phase space $\mathcal{S}(\mathbf{x}, \mathbf{p}; \gamma)$ defined by Eq. (6), the value of the action sum is conserved during the evolution (Eq. (21)), i.e.,

$$e_t^{(1)} + e_t^{(2)} = 1 + F\gamma = 1 + 2\gamma . \tag{24}$$

Throughout this paper, we use $\mod[\alpha, 2\pi]$ to represent the remainder after dividing an angle $\alpha$ by $2\pi$, which lies in range $[0, 2\pi)$. Define the difference between the two angle variables

$$\theta^d = \mod\left[\theta^{(2)} - \theta^{(1)}, 2\pi\right] , \tag{25}$$

whose initial value is $\theta_0^d$. Substitution of Eqs. (22)-(23) in Eq. (21) produces the analytical solution of $\mathbf{g}_t \equiv \mathbf{x}_t + i\mathbf{p}_t$. It is straightforward to show

$$\begin{aligned} e_t^{(1)} &= (1 + F\gamma)\sin^2 \xi + e_0^{(1)} \cos(2\xi) \\ &\quad + \sqrt{e_0^{(1)}(1 + F\gamma - e_0^{(1)})} \sin(2\xi) \cos(\varphi - \psi + \theta_0^d) \end{aligned} \tag{26}$$

Eq. (24) for the action sum and Eq. (26) indicate that action variables $\{e_t^{(1)}, e_t^{(2)}\}$ depend on only $\theta_0^d$ rather than on two initial angle values $\{\theta_0^{(1)}, \theta_0^{(2)}\}$. That is, $\mathbf{e}_t = \mathbf{e}_t(\mathbf{e}_0, \theta_0^d)$. In addition, Eq. (26) suggests that $\mathbf{e}_t(\mathbf{e}_0, \theta_0^d)$ is periodic with respect to the initial value $\theta_0^d$, i.e.,

$$\mathbf{e}_t\left(\mathbf{e}_0,\theta_0^d\right) \equiv \mathbf{e}_t\left(\mathbf{e}_0,\theta_0^d + 2n\pi\right) \quad , \tag{27}$$

where $n$ is an integer.

Define the scaled action difference

$$y = \frac{e^{(1)} - e^{(2)}}{2(1+F\gamma)} = \frac{e^{(1)}}{1+F\gamma} - \frac{1}{2} \quad . \tag{28}$$

The value of $y$ lies in region $[-1/2, 1/2]$. Substitution of Eq. (26) into Eq. (28) for time $t$ yields

$$y_t = y_0 \cos(2\xi) + \sqrt{\frac{1}{4} - (y_0)^2} \sin(2\xi) \cos(\varphi - \psi + \theta_0^d) \quad , \tag{29}$$

where $y_0 = \frac{e_0^{(1)}}{1+F\gamma} - \frac{1}{2}$ for time 0. When $\sin(2\xi(t)) = 0$, $y_t = y_0$ or $y_t = -y_0$, so that $y_t(y_0, \theta_0^d)$ is irrelevant to $\theta_0^d$ for all $y_0 \in [-1/2, 1/2]$.

### III. A Novel Class of Isomorphic Representations for Population Dynamics of the Two-State System

#### III-A. Expression of the Population-Population Correlation Function

We aim to propose a class of novel phase space representations for the exact population-population correlation function of TSS. The essential idea lies in the generalization of the phase space mapping formalism of Eq. (17). In this paper, we introduce a generalized weight function $\bar{w}(\mathbf{x}_0, \mathbf{p}_0; \mathbf{x}_t, \mathbf{p}_t)$ that depends both on the values of mapping phase space variables at time 0 and time $t$, and a time-dependent normalization factor $\bar{C}_{nn,mm}(t)$. It leads to the new phase space representation of the population-population correlation function of TSS,

$$\begin{cases} \mathrm{Tr}_e\left[|n\rangle\langle n|e^{i\hat{H}t}|m\rangle\langle m|e^{-i\hat{H}t}\right] \mapsto \left(\bar{C}_{nn,mm}(t)\right)^{-1} p_{n\to m}(t) \\ p_{n\to m}(t) = \int_{\mathcal{S}(\mathbf{x}_0,\mathbf{p}_0;\gamma)} F d\mathbf{x}_0 d\mathbf{p}_0 \bar{w}(\mathbf{x}_0,\mathbf{p}_0;\mathbf{x}_t,\mathbf{p}_t) K_{nn}(\mathbf{x}_0,\mathbf{p}_0) K_{mm}(\mathbf{x}_t,\mathbf{p}_t) \,, \\ \bar{C}_{nn,mm}(t) = \sum_{k=1}^{2} p_{n\to k}(t) \end{cases} \quad (30)$$

where the phase space counterpart of pure state $|1\rangle\langle 1|$ is the Heaviside step function

$$K_{11}(\mathbf{x},\mathbf{p}) = K_{11}(y(\mathbf{x},\mathbf{p})) = h(y(\mathbf{x},\mathbf{p})) \,, \quad (31)$$

and that of pure state $|2\rangle\langle 2|$ is

$$K_{22}(\mathbf{x},\mathbf{p}) = K_{22}(y(\mathbf{x},\mathbf{p})) = h(-y(\mathbf{x},\mathbf{p})). \quad (32)$$

In Eq. (30), the generalized weight function $\bar{w}(\mathbf{x}_0,\mathbf{p}_0;\mathbf{x}_t,\mathbf{p}_t)$ is a function of both $y_0(\mathbf{x}_0,\mathbf{p}_0)$ and $y_t(\mathbf{x}_t,\mathbf{p}_t)$:

$$\bar{w}(\mathbf{x}_0,\mathbf{p}_0;\mathbf{x}_t,\mathbf{p}_t) = \bar{w}(y_0(\mathbf{x}_0,\mathbf{p}_0); y_t(\mathbf{x}_t,\mathbf{p}_t)) \\ = \begin{cases} f(|y_0|) & \text{if } |y_0| < |y_t|, \\ f(|y_t|) & \text{if } |y_0| \geq |y_t| \end{cases}, \quad (33)$$

where $f(y)$ is a function to be determined. We *demand* that $f(y)$ is bounded and non-negative for $y \in [0, 1/2]$. Equation (33) offers a generalized time-dependent weight function. As will be discussed in detail in Section IV, the SQC triangle window functions lead to a special case of the generalized time-dependent weight function. When $\mathbf{g}_t = \mathbf{U}(t)\mathbf{g}_0$ follows Eq. (21), mapping kernel $\hat{K}(\mathbf{x},\mathbf{p})$ is covariant[56] if

$$\hat{K}(\mathbf{x}_t,\mathbf{p}_t) = \mathbf{U}(t)\hat{K}(\mathbf{x}_0,\mathbf{p}_0)\mathbf{U}^{\dagger}(t) \quad. \quad (34)$$

When the diagonal terms of $\hat{K}(\mathbf{x},\mathbf{p})$ include $K_{11}(\mathbf{x},\mathbf{p})$ and $K_{22}(\mathbf{x},\mathbf{p})$ of Eqs. (31)-(32), and off-diagonal terms of $\hat{K}(\mathbf{x},\mathbf{p})$ are those defined in Eq. (8), the condition of covariance, Eq. (34), does not hold. Non-covariant phase space functions are then utilized in the mapping formalism of Eqs. (30)-(33).

In the next subsection we will demonstrate that for the mapping formalism of Eqs. (30)-(33), the time-dependent normalization factor $\bar{C}_{nn,mm}(t)$ depends only on $\xi(t)$, and for all $n$ and $m$, $\bar{C}_{nn,mm}(t)$ are the same. Thus, for convenience, we define

$$\Xi(\xi) \equiv \bar{C}_{nn,mm}(t) \tag{35}$$

when $\xi$ is viewed as an independent variable. We prove a theorem of the exact population dynamics of TSS:

**Theorem.** If $\Xi(\xi)$ is a bounded smooth function for $\xi \in [0, \pi/2]$, satisfying $\Xi\left(\dfrac{\pi}{2} - \xi\right) = \Xi(\xi)$ and $\Xi''(0) = 2\Xi(0) = 2$, then the function $f(y)$ satisfying

$$\begin{cases} B(z) = z\sqrt{1-z^2}\,\dfrac{\mathrm{d}}{\mathrm{d}z}\left(z^2 \Xi(\arcsin z)\right) \\ f(y) = \displaystyle\int_0^{2y} \dfrac{\mathrm{d}z}{\sqrt{(2y)^2 - z^2}}\,\dfrac{\mathrm{d}B(z)}{\mathrm{d}z} \end{cases} \tag{36}$$

is bounded on $y \in [0, 1/2]$. Substitution of the expression of $f(y)$ of Eq. (36) into the generalized weight function of Eq. (33) proves that the population-population correlation function of the mapping formalism of Eqs. (30)-(33) of TSS is exact, i.e.,

$$\begin{aligned}
|U_{mn}(t)|^2 &\equiv \mathrm{Tr}_e\left[|n\rangle\langle n|e^{i\hat{H}t}|m\rangle\langle m|e^{-i\hat{H}t}\right] \\
&= \left(\bar{C}_{nn,mm}(t)\right)^{-1} p_{n\to m}(t) \\
&= \left(\bar{C}_{nn,mm}(t)\right)^{-1} \int_{\mathcal{S}(\mathbf{x}_0,\mathbf{p}_0;\gamma)} F\mathrm{d}\mathbf{x}_0\mathrm{d}\mathbf{p}_0\,\bar{w}(\mathbf{x}_0,\mathbf{p}_0;\mathbf{x}_t,\mathbf{p}_t) K_{nn}(\mathbf{x}_0,\mathbf{p}_0) K_{mm}(\mathbf{x}_t,\mathbf{p}_t)
\end{aligned} \tag{37}$$

In addition, we demand $f(y) \geq 0$ for satisfying the non-negative property of the mapping formalism Eqs. (30)-(33).

As shown in Appendix B, when the generalized weight function $\bar{w}(\mathbf{x}_0, \mathbf{p}_0; \mathbf{x}_t, \mathbf{p}_t)$

takes the form of Eq. (33), the expression of $p_{n \to m}(t)$ with variables $(y_0, \theta_0^d)$ reads

$$p_{n \to m}(t) = \frac{1}{\pi} \int_{-1/2}^{1/2} dy_0 \int_0^{2\pi} d\theta_0^d \, \overline{w}(y_0; y_t) K_{nn}(y_0) K_{mm}(y_t) \ . \tag{38}$$

As discussed in Appendix C, when $\sin(2\xi(t)) = 0$, the property of Eq. (38) implies that the phase space mapping formalism of Eqs. (30)-(33) *naturally* satisfies the equality of Eq. (37), the condition of the exact population dynamics of TSS. The more *non-trivial* case $\sin(2\xi(t)) \neq 0$ is then our focal point, for which we will prove that the phase space formalism of Eqs. (30)-(33) with Eq. (36) also satisfies the same condition of Eq. (37).

In Subsection III-B, we will show the crucial symmetry of the time-dependent normalization factor related to the exact population dynamics of TSS. In Subsection III-C, we will prove that the $f(y)$ defined by Eq. (36) satisfies this symmetry, which leads to the exact population-population correlation function of TSS.

**III-B.  Symmetry in Dynamics of the Two-State System Related to the Exact Population Dynamics**

When the general (non-trivial) case $\sin(2\xi(t)) \neq 0$ is considered, we change the variables of Eq. (38) from $(y_0, \theta_0^d)$ to $(y_0, y_t)$, in order to conveniently utilize the symmetry of $\overline{w}(\mathbf{x}_0, \mathbf{p}_0; \mathbf{x}_t, \mathbf{p}_t)$ with the form of Eq. (33). In Appendix D, we show

$$\begin{aligned}&\frac{1}{\pi}\int_{-1/2}^{1/2} dy_0 \int_0^{2\pi} d\theta_0^d \, \overline{w}(y_0; y_t) K_{nn}(y_0) K_{mm}(y_t) \\ &= \frac{2}{\pi} \int_\Omega dy_0 dy_t \, \frac{\overline{w}(y_0; y_t) K_{nn}(y_0) K_{mm}(y_t)}{\sqrt{\mathcal{E}(y_0, y_t)}}\end{aligned}, \tag{39}$$

where $\Omega$ is an ellipse-shaped domain defined as

$$\Omega := \{(y_0, y_t) \mid \mathcal{E}(y_0, y_t) > 0\} \ , \tag{40}$$

where

$$\mathcal{E}(y_0, y_t) = \frac{\sin^2(2\xi)}{4} - (y_0)^2 - (y_t)^2 + 2y_0 y_t \cos(2\xi) \ . \tag{41}$$

Figure 2 illustrates the integration domain of $(y_0, \theta_0^d)$ and that of $(y_0, y_t)$ in Eq. (39).

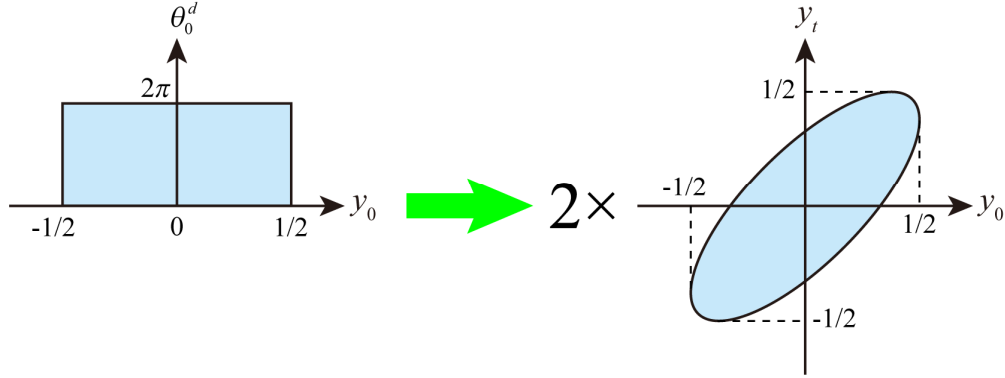

**Figure 2.** The integration domain of $(y_0, \theta_0^d)$ and that of $(y_0, y_t)$ in Eq. (39). The integration domain of $(y_0, \theta_0^d)$ is the rectangle $(-1/2, 1/2) \otimes [0, 2\pi)$, and the integration domain of $(y_0, y_t)$ is the ellipse $\Omega$. The expression of $\Omega$ is defined by Eqs. (40)-(41). The value of integration of function $\sigma(y_0; y_t)$ of variables $(y_0, \theta_0^d)$ on $(-1/2, 1/2) \otimes [0, 2\pi)$ is twice of the value of integration of $\sigma(y_0; y_t) / \sqrt{\mathcal{E}(y_0, y_t)}$ of variables $(y_0, y_t)$ on $\Omega$.

Both the integrand function $\bar{w}(y_0; y_t) / \sqrt{\mathcal{E}(y_0, y_t)}$ and the integration domain $\Omega$ of Eq. (39) have the central symmetry, i.e., they keep the same while simultaneously changing the signs of $y_0$ and $y_t$. As shown in Appendix E, it is straightforward from the central symmetry to verify

$$\begin{aligned} p_{1 \to 1}(t) &= p_{2 \to 2}(t) \\ p_{1 \to 2}(t) &= p_{2 \to 1}(t) \end{aligned} \tag{42}$$

so that for all $n$ and $m$, $\bar{C}_{nn,mm}(t)$ are the same. Equation (42) holds for only TSS. We use Eq. (38) and the integral identity Eq. (39) to simplify $p_{n\to m}(t)$ defined in Eq. (30). It is straightforward to show

$$p_{1\to 1}(t) = \int_{\mathcal{S}(\mathbf{x}_0,\mathbf{p}_0;\gamma)} F d\mathbf{x}_0 d\mathbf{p}_0 \bar{w}(\mathbf{x}_0,\mathbf{p}_0;\mathbf{x}_t,\mathbf{p}_t) K_{11}(\mathbf{x}_0,\mathbf{p}_0) K_{11}(\mathbf{x}_t,\mathbf{p}_t)$$
$$= \frac{2}{\pi} \int_{\Omega_{1\to 1}} dy_0 dy_t \frac{\bar{w}(y_0;y_t)}{\sqrt{\mathcal{E}(y_0,y_t)}} \tag{43}$$

where $\Omega_{1\to 1} = \{(y_0,y_t) | (y_0,y_t) \in \Omega, y_0 > 0, y_t > 0\}$, and

$$p_{1\to 2}(t) = \int_{\mathcal{S}(\mathbf{x}_0,\mathbf{p}_0;\gamma)} F d\mathbf{x}_0 d\mathbf{p}_0 \bar{w}(\mathbf{x}_0,\mathbf{p}_0;\mathbf{x}_t,\mathbf{p}_t) K_{11}(\mathbf{x}_0,\mathbf{p}_0) K_{22}(\mathbf{x}_t,\mathbf{p}_t)$$
$$= \frac{2}{\pi} \int_{\Omega_{1\to 2}} dy_0 dy_t \frac{\bar{w}(y_0;y_t)}{\sqrt{\mathcal{E}(y_0,y_t)}} \tag{44}$$

where $\Omega_{1\to 2} = \{(y_0,y_t) | (y_0,y_t) \in \Omega, y_0 > 0, y_t < 0\}$. The expressions of $\bar{w}(y_0;y_t)$ and $\mathcal{E}(y_0,y_t)$ indicate that lines $y_t = \pm y_0$ are the two axes of symmetry of $\bar{w}(y_0;y_t)/\sqrt{\mathcal{E}(y_0,y_t)}$. Thus, the minimum integration domains, with respect to the symmetry, include

$$\Omega_{1\to 1}^{\text{rdn}} = \{(y_0,y_t) | (y_0,y_t) \in \Omega, y_0 > 0, 0 < y_t < y_0\} \tag{45}$$

and

$$\Omega_{1\to 2}^{\text{rdn}} = \{(y_0,y_t) | (y_0,y_t) \in \Omega, y_0 > 0, -y_0 < y_t < 0\} \tag{46}$$

Figure 3 illustrates the integration domains, $\Omega_{1\to 1}$, $\Omega_{1\to 2}$, $\Omega_{1\to 1}^{\text{rdn}}$, and $\Omega_{1\to 2}^{\text{rdn}}$.

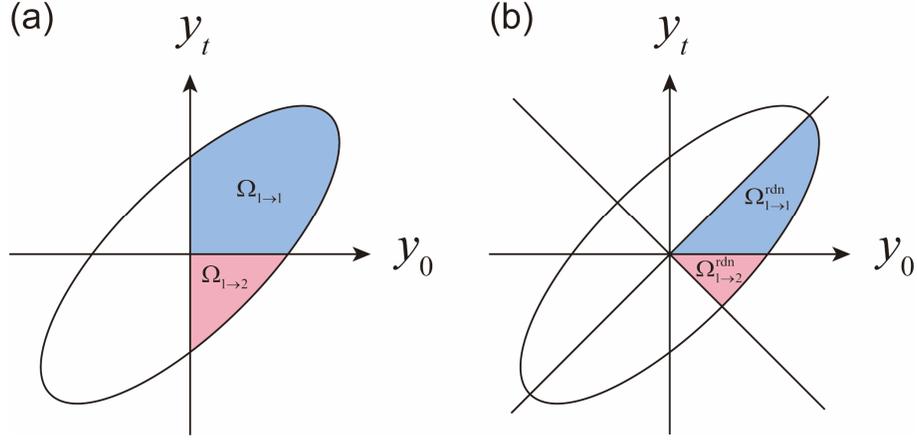

**Figure 3.** Panel (a) depicts the domains, $\Omega_{1\to 1}$ and $\Omega_{1\to 2}$. Panel (b) illustrates the regions $\Omega_{1\to 1}^{rdn}$ and $\Omega_{1\to 2}^{rdn}$, the minimum integration domains with respect to the symmetry. The two additional lines indicating the symmetries in Panel (b) are $y_t = \pm y_0$.

On domains $\Omega_{1\to 1}^{rdn}$ and $\Omega_{1\to 2}^{rdn}$, we transform the coordinate system of $(y_0, y_t)$ into the following polar coordinate system of $(s, T)$, which is more convenient to use for the ellipse, $\Omega$,

$$\begin{cases} y_0 = \dfrac{\sin(s)\sin(T+\xi(t))}{2} \\ y_t = \dfrac{\sin(s)\sin(T-\xi(t))}{2} \end{cases}, \qquad (47)$$

where $s \in [0, \pi/2)$ and $T \in [0, 2\pi)$. By using identical transformations among trigonometric functions, it is straightforward to derive $|\partial(y_0, y_t)/\partial(s,T)| = \sin(2s)\sin(2\xi(t))/8$. Substitution of Eq. (47) into Eq. (41) yields $\mathcal{E}(y_0, y_t) = \sin^2(2\xi(t))\cos^2(s)/4$. Utilizing the symmetry of the generalized weight function of Eq. (33), and using the expressions of $|\partial(y_0, y_t)/\partial(s,T)|$ and

$\mathcal{E}(y_0, y_t)$, we express the integrals of Eq. (43) and Eq. (44) with variables $(s, T)$:

$$p_{1 \to 1}(t) = \frac{2}{\pi} \int_{\Omega_{1 \to 1}} dy_0 dy_t \frac{\overline{w}(y_0; y_t)}{\sqrt{\mathcal{E}(y_0, y_t)}}$$

$$= \frac{4}{\pi} \int_{\Omega_{1 \to 1}^{\text{rdn}}} dy_0 dy_t \frac{f(y_t)}{\sqrt{\mathcal{E}(y_0, y_t)}} \qquad (48)$$

$$= \frac{2}{\pi} \int_0^{\pi/2} ds \int_{\xi(t)}^{\pi/2} dT \sin(s) f\left(\sin(s)\sin(T - \xi(t))/2\right)$$

$$\stackrel{\tau = T - \xi(t)}{=} \frac{2}{\pi} \int_0^{\pi/2} ds \int_0^{\pi/2 - \xi(t)} d\tau \sin(s) f\left(\sin(s)\sin(\tau)/2\right)$$

and

$$p_{1 \to 2}(t) = \frac{2}{\pi} \int_{\Omega_{1 \to 2}} dy_0 dy_t \frac{\overline{w}(y_0; y_t)}{\sqrt{\mathcal{E}(y_0, y_t)}}$$

$$= \frac{4}{\pi} \int_{\Omega_{1 \to 2}^{\text{rdn}}} dy_0 dy_t \frac{f(-y_t)}{\sqrt{\mathcal{E}(y_0, y_t)}} \qquad (49)$$

$$= \frac{2}{\pi} \int_0^{\pi/2} ds \int_0^{\xi(t)} dT \sin(s) f\left(-\sin(s)\sin(T - \xi(t))/2\right)$$

$$\stackrel{\tau = \xi(t) - T}{=} \frac{2}{\pi} \int_0^{\pi/2} ds \int_0^{\xi(t)} d\tau \sin(s) f\left(\sin(s)\sin(\tau)/2\right)$$

It is evident from Eqs. (48)-(49) that $\overline{C}_{nn,mm}(t)$ depends only on $\xi(t)$. The condition of the exact population dynamics of TSS, the equality of the first line of Eq. (37), i.e.,

$$|U_{mn}(t)|^2 = p_{n \to m}(t) / \overline{C}_{nn,mm}(t) , \qquad (50)$$

holds if and only if

$$\frac{p_{1 \to 1}(t)}{|U_{11}(t)|^2} = \frac{p_{1 \to 2}(t)}{|U_{21}(t)|^2} \qquad (51)$$

for the *non-trivial* case $\sin(2\xi(t)) \neq 0$. Substitution of Eq. (22) into Eq. (51) yields

$$\frac{p_{1 \to 1}(t)}{\cos^2(\xi(t))} = \frac{p_{1 \to 2}(t)}{\sin^2(\xi(t))} = \overline{C}_{nn,mm}(t) \quad . \qquad (52)$$

Substituting Eqs. (48)-(49) into Eq. (52), and utilizing $\Xi(\xi) \equiv \overline{C}_{nn,mm}(t)$ defined in

Eq. (35), we obtain the important axial symmetry

$$\Xi(\xi) = \frac{2}{\pi}\int_0^{\pi/2} ds \int_0^{\xi} d\tau \frac{\sin s}{\sin^2 \xi} f(\sin(s)\sin(\tau)/2)$$
$$= \frac{2}{\pi}\int_0^{\pi/2} ds \int_0^{\pi/2-\xi} d\tau \frac{\sin s}{\cos^2 \xi} f(\sin(s)\sin(\tau)/2) \quad (53)$$
$$= \Xi\left(\frac{\pi}{2} - \xi\right)$$

The axis of symmetry is $\xi = \pi/4$. The axial symmetry of Eq. (53) guarantees that for the mapping formalism Eqs. (30)-(33), the equality of Eq. (37), the condition of the exact population dynamics of TSS, holds for the general non-trivial case $\sin(2\xi(t)) \neq 0$. Because $\xi$ lies in region $[0, \pi/2]$ as discussed around Eqs. (95)-(96) of Appendix A, we always have $\sin^2 \xi \neq 0$ when the case $\sin(2\xi(t)) \neq 0$ is considered. Multiplying $\pi \sin^2 \xi$ in both sides of the first equation of Eq. (53) leads to the integral equation

$$\pi \Xi(\xi) \sin^2(\xi)/2 = \int_0^{\pi/2} ds \int_0^{\xi} d\tau \sin(s) f(\sin(s)\sin(\tau)/2) \quad . \quad (54)$$

**III-C. Proof of the Exact Population Dynamics of the Novel Class of Isomorphic Representations**

The axial symmetry $\Xi\left(\frac{\pi}{2} - \xi\right) = \Xi(\xi)$ serves as the crucial point for constructing a novel class of isomorphic representations of *the exact population dynamics* of TSS. By choosing function $\Xi(\xi)$ satisfying the axial symmetry $\Xi\left(\frac{\pi}{2} - \xi\right) = \Xi(\xi)$ of Eq. (53), we obtain function $f(y)$ from the solution to the integral equation of Eq. (54). We *demand* that $f(\sin(s)\sin(\tau)/2)$ is a non-negative bounded function on the region, $\sin(s)\sin(\tau) \in [0,1]$ (which is irrelevant to the proof of the exact population-population correlation function, Eq. (37)). Because the physical meaning of $\Xi(\xi)$ is the time-dependent normalization factor, $\Xi(\xi)$ is a smooth function that is bounded on the interval $\xi \in [0, \pi/2]$. Since $f(y)$ is bounded, when $\xi$ approaches zero, both the left hand side (LHS) and right hand side (RHS) of Eq. (54) approach zero as well. We can thus solve Eq. (54) by differentiating both sides of the equation over $\xi$, i.e., obtain the solution to

$$\int_0^{\pi/2} ds \sin(s) f\left(\sin(s)\sin(\xi)/2\right) = \frac{\pi}{2} \frac{d}{d\xi}\left(\Xi(\xi)\sin^2 \xi\right) \qquad (55)$$

instead. By making changes of variables $x = \sin(s)\sin(\xi)$ and $z = \sin\xi$, Eq. (55) becomes a variant of *Abel equation* that can be exactly solved[51],

$$\int_0^z \frac{xf(x/2)dx}{\sqrt{z^2 - x^2}} = \frac{\pi B(z)}{2}, \qquad (56)$$

where

$$B(z) = z\sqrt{1-z^2}\frac{d}{dz}\left(z^2 \Xi(\arcsin z)\right). \qquad (57)$$

We follow Ref. [51] to show the procedure for achieving the solution to Eq. (56). Let $u = z^2$ and $v = x^2$, Eq. (56) becomes

$$\int_0^u \frac{f\left(\sqrt{v}/2\right)dv}{\sqrt{u-v}} = \pi B\left(\sqrt{u}\right). \qquad (58)$$

Integrating both sides of Eq. (58) over $u$ from 0 to $\tilde{v}$ after multiplying $1/\sqrt{\tilde{v}-u}$, the equation becomes

$$\int_0^{\tilde{v}} \frac{du}{\sqrt{\tilde{v}-u}} \int_0^u \frac{dv}{\sqrt{u-v}} f\left(\sqrt{v}/2\right) = \pi \int_0^{\tilde{v}} \frac{du}{\sqrt{\tilde{v}-u}} B\left(\sqrt{u}\right). \qquad (59)$$

Utilizing

$$\int_{v}^{\tilde{v}} \frac{du}{\sqrt{\tilde{v}-u}\sqrt{u-v}} = \pi \tag{60}$$

and that $f(\sqrt{v}/2)$ is continuous, the LHS of Eq. (59) becomes

$$\int_{0}^{\tilde{v}} \frac{du}{\sqrt{\tilde{v}-u}} \int_{0}^{u} \frac{dv}{\sqrt{u-v}} f(\sqrt{v}/2)$$
$$= \int_{0}^{\tilde{v}} f(\sqrt{v}/2) dv \int_{v}^{\tilde{v}} \frac{du}{\sqrt{\tilde{v}-u}\sqrt{u-v}} \tag{61}$$
$$= \pi \int_{0}^{\tilde{v}} f(\sqrt{v}/2) dv$$

Substituting Eq. (61) into Eq. (59) and differentiating both sides over $\tilde{v}$, we transform Eq. (59) to

$$f(\sqrt{\tilde{v}}/2) = \frac{d}{d\tilde{v}} \int_{0}^{\tilde{v}} \frac{B(\sqrt{u})}{\sqrt{\tilde{v}-u}} du \quad . \tag{62}$$

Utilizing $B(0) = 0$, and letting $\tilde{u} = \tilde{v} - u$, we simplify Eq. (62) below,

$$\frac{d}{d\tilde{v}} \int_{0}^{\tilde{v}} \frac{B(\sqrt{u})}{\sqrt{\tilde{v}-u}} du$$
$$= \frac{d}{d\tilde{v}} \int_{0}^{\tilde{v}} \frac{B(\sqrt{\tilde{v}-\tilde{u}})}{\sqrt{\tilde{u}}} d\tilde{u}$$
$$= \frac{B(0)}{\sqrt{\tilde{v}}} + \int_{0}^{\tilde{v}} \frac{1}{\sqrt{\tilde{u}}} \frac{\partial B(\sqrt{\tilde{v}-\tilde{u}})}{\partial \tilde{v}} d\tilde{u} \quad . \tag{63}$$
$$= 0 - \int_{0}^{\tilde{v}} \frac{1}{\sqrt{\tilde{u}}} \frac{\partial B(\sqrt{\tilde{v}-\tilde{u}})}{\partial \tilde{u}} d\tilde{u}$$
$$= \int_{0}^{\tilde{v}} \frac{1}{\sqrt{\tilde{v}-u}} \frac{\partial B(\sqrt{u})}{\partial u} du$$

By reassigning $y$ to be $\sqrt{\tilde{v}}/2$ and letting $z = \sqrt{u}$, we obtain the solution to Eq. (56),

$$f(y) = \int_{0}^{2y} \frac{dz}{\sqrt{(2y)^2 - z^2}} \frac{dB(z)}{dz} \quad . \tag{64}$$

As shown in Appendix F, $\lim_{y \to 0+} f(y) = 0$, and $f(y)$ is bounded for $y \in [0, 1/2]$

only if $\Xi''(0) = 2\Xi(0) = 2$. As a consequence of $\Xi''(0) \neq 0$, when the off-diagonal terms of the Hamiltonian matrix $\mathbf{H}$ are not zero, the time-dependent normalization factors are not constants. Equation (64) satisfies Eq. (37), the condition of reproducing the exact population dynamics of TSS with the mapping formalism of Eqs. (30)-(33). We additionally demand $f(y) \geq 0$.

Tidying up Eqs. (53), (57) and (64), we derive the following novel class of isomorphic representations of the *exact* population-population correlation function of TSS,

$$\begin{cases} \mathrm{Tr}_e\left[|n\rangle\langle n|e^{i\hat{H}t}|m\rangle\langle m|e^{-i\hat{H}t}\right] \mapsto \left(\overline{C}_{nn,mm}(t)\right)^{-1} p_{n\to m}(t) \\ p_{n\to m}(t) = \int_{S(\mathbf{x}_0,\mathbf{p}_0;\gamma)} F \mathrm{d}\mathbf{x}_0 \mathrm{d}\mathbf{p}_0 \overline{w}(\mathbf{x}_0,\mathbf{p}_0;\mathbf{x}_t,\mathbf{p}_t) K_{nn}(\mathbf{x}_0,\mathbf{p}_0) K_{mm}(\mathbf{x}_t,\mathbf{p}_t) \\ \overline{C}_{nn,mm}(t) = \sum_{k=1}^{2} p_{n\to k}(t) \\ y = \dfrac{1}{2(1+F\gamma)}\left(\left(x^{(1)}\right)^2 + \left(p^{(1)}\right)^2\right) - 1/2 \\ K_{11}(\mathbf{x},\mathbf{p}) = h(y) \\ K_{22}(\mathbf{x},\mathbf{p}) = h(-y) \\ \overline{w}(\mathbf{x}_0,\mathbf{p}_0;\mathbf{x}_t,\mathbf{p}_t) = \begin{cases} f(|y_0|) & \text{if } |y_0| < |y_t|, \\ f(|y_t|) & \text{if } |y_0| \geq |y_t| \end{cases} \end{cases} \quad (65)$$

with

$$\begin{cases} B(z) = z\sqrt{1-z^2}\,\dfrac{\mathrm{d}}{\mathrm{d}z}\left(z^2 \Xi(\arcsin z)\right) \\ f(y) = \displaystyle\int_0^{2y} \dfrac{\mathrm{d}z}{\sqrt{(2y)^2 - z^2}} \dfrac{\mathrm{d}B(z)}{\mathrm{d}z} \end{cases}, \quad (66)$$

where $\Xi\left(\dfrac{\pi}{2} - \xi\right) = \Xi(\xi)$ and $\Xi''(0) = 2\Xi(0) = 2$. In addition, we require $f(y) \geq 0$ for the non-negative property of the mapping formalism of Eqs. (30)-(33).

In the novel isomorphism class defined by Eqs. (65)-(66), all phase space functions, $K_{nn}(\mathbf{x}_0, \mathbf{p}_0)$ and $K_{mm}(\mathbf{x}_t, \mathbf{p}_t)$, and the time-dependent weight function, $\overline{w}(\mathbf{x}_0, \mathbf{p}_0; \mathbf{x}_t, \mathbf{p}_t)$, are positive semi-definite. Because $\overline{w}(\mathbf{x}_0, \mathbf{p}_0; \mathbf{x}_t, \mathbf{p}_t) K_{nn}(\mathbf{x}_0, \mathbf{p}_0) K_{mm}(\mathbf{x}_t, \mathbf{p}_t)$, the integrand function of the integral over CPS in the RHS of the second equation of Eq. (65), is non-negative, $p_{n \to m}(t)$ as well as the time-dependent normalization factor, $\overline{C}_{nn,mm}(t)$, remains positive semi-definite at any time. It is then evident that *the contribution of each trajectory* (where the coordinate-momentum variables $(\mathbf{x}_t, \mathbf{p}_t)$ at time $t$ are produced by the EOMs of Eq. (20) with the initial condition $(\mathbf{x}_0, \mathbf{p}_0)$ at time 0 on CPS) to the integral expression of the population-population correlation function is rigorously *positive semi-definite*.

In Figure 4, we plot two new examples of $f(y)$ of the function family Eq. (66) and their corresponding time-dependent normalization factors $\Xi(\xi)$, with comparison to their counterparts derived from the triangle window functions of the SQC approach of Ref. [28]. The two new functions are labeled as Case 1 and Case 2. In the next section, we will prove that the triangle window functions of the SQC approach lead to a special case of the novel class of isomorphic representations of TSS. The analytical expressions of the functions shown in Figure 4 are listed in Appendix G. These example functions for $f(y)$ shown in Figure 4, which are strictly *positive semi-definite*, lead to isomorphic representations of the exact population dynamics of TSS.

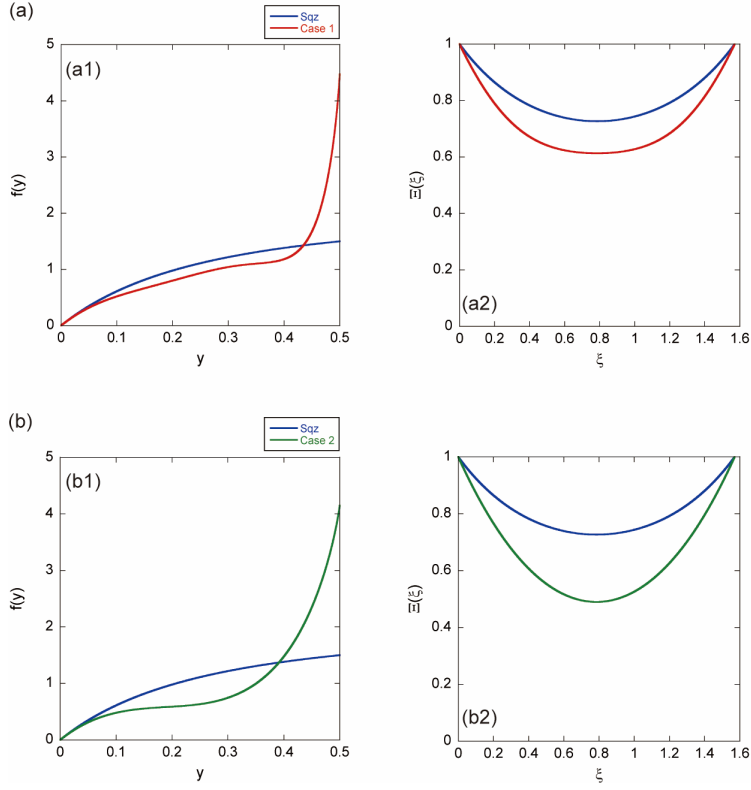

**Figure 4.** Panel (a1) depicts an example of the function family Eq. (66) of $f(y)$, which is denoted as Case 1, as well as the particular $f(y)$ in the squeezed (sqz) SQC-TWF approach. Panel (b1) does so for another example of $f(y)$ denoted as Case 2. Panel (a2) depicts the time-dependent normalization factor $\Xi(\xi) \equiv \overline{C}_{nn,mm}(t)$ corresponding to Case 1. Panel (b2) does so for Case 2. The explicit formulas for $f(y)$ in Panels (a1) and (b1), and those for $\Xi(\xi)$ in Panels (a2) and (b2), respectively, are listed in Appendix G. In all panels, the blue, red, and green lines stand for squeezed SQC-TWF, Case 1, and Case 2, respectively.

## IV. Case Study: The Symmetrical Quasi-Classical (SQC) Triangle Window Functions of TSS

### IV-A. Original Expression of the Population-Population Correlation Function

**of the SQC-TWF Approach**

The TWF approach in the SQC method of Cotton and Miller uses triangle window functions of the action space to express population dynamics[28, 30]. For the $F=2$ case, the population-population correlation function reads

$$\text{Tr}_e\left[|n\rangle\langle n|e^{i\hat{H}t}|m\rangle\langle m|e^{-i\hat{H}t}\right] \mapsto \left(\bar{C}_{nn,mm}^{\text{SQC}}(t)\right)^{-1} p_{n\to m}^{\text{SQC}}(t) \quad, \tag{67}$$

where

$$p_{n\to m}^{\text{SQC}}(t) = \int \frac{2\mathrm{d}\mathbf{x}_0 \mathrm{d}\mathbf{p}_0}{(2\pi)^2} K_{nn}^{\text{SQC}}(\mathbf{x}_0,\mathbf{p}_0) K_{mm}^{\text{bin}}(\mathbf{x}_t,\mathbf{p}_t), \tag{68}$$

and the time-dependent normalization factor is defined as

$$\bar{C}_{nn,mm}^{\text{SQC}}(t) = \sum_{k=1}^{2} \int \frac{2\mathrm{d}\mathbf{x}_0 \mathrm{d}\mathbf{p}_0}{(2\pi)^2} K_{nn}^{\text{SQC}}(\mathbf{x}_0,\mathbf{p}_0) K_{kk}^{\text{bin}}(\mathbf{x}_t,\mathbf{p}_t), \tag{69}$$

of which the initial value $\bar{C}_{nn,mm}^{\text{SQC}}(0)$ is 1. The EOMs of $(\mathbf{x}_t,\mathbf{p}_t)$ are isomorphic to the TDSE, where $(\mathbf{x}_0,\mathbf{p}_0)$ is the initial condition.

The phase space function, $K_{nn}^{\text{SQC}}(\mathbf{x},\mathbf{p})$, is defined in the sampling region of phase space variables corresponding to the $n$-th state, which is

$$K_{nn}^{\text{SQC}}(\mathbf{x},\mathbf{p}) = \begin{cases} 1 & \text{if } (\mathbf{x},\mathbf{p}) \in \mathcal{M}_n \\ 0 & \text{otherwise} \end{cases} \tag{70}$$

where $\mathcal{M}_n$ is the manifold of phase space points of triangle window form,

$$\mathcal{M}_n(\mathbf{x},\mathbf{p}) : \left\{(\mathbf{x},\mathbf{p}) \middle| \begin{array}{l} 1 \leq \frac{1}{2}\left(\left(x^{(n)}\right)^2 + \left(p^{(n)}\right)^2\right) \leq 2; \\ \frac{1}{2}\left(\left(x^{(n'\neq n)}\right)^2 + \left(p^{(n'\neq n)}\right)^2 + \left(x^{(n)}\right)^2 + \left(p^{(n)}\right)^2\right) \leq 2 \end{array}\right\}. \tag{71}$$

When $n=1$, then $n'=2$, and *vice versa*. The other phase space function, $K_{mm}^{\text{bin}}(\mathbf{x},\mathbf{p})$ for the $m$-th state is

$$K_{mm}^{\text{bin}}(\mathbf{x},\mathbf{p}) = \begin{cases} 1 & \text{if } (\mathbf{x},\mathbf{p}) \in \mathcal{M}_m^{\text{bin}} \\ 0 & \text{otherwise} \end{cases}, \tag{72}$$

where

$$\mathcal{M}_m^{\text{bin}}(\mathbf{x},\mathbf{p}) : \left\{ (\mathbf{x},\mathbf{p}) \middle| \begin{array}{l} 1 \leq \frac{1}{2}\left( \left(x^{(m)}\right)^2 + \left(p^{(m)}\right)^2 \right); \\ \frac{1}{2}\left( \left(x^{(m'\neq m)}\right)^2 + \left(p^{(m'\neq m)}\right)^2 \right) \leq 1 \end{array} \right\}. \qquad (73)$$

The positive semi-definite phase space functions, $K_{nn}^{\text{SQC}}(\mathbf{x},\mathbf{p})$ and $K_{mm}^{\text{bin}}(\mathbf{x},\mathbf{p})$, can alternatively be expressed with action-angle variables $\{\mathbf{e},\boldsymbol{\theta}\}$ as [28, 30],

$$\begin{aligned}
K_{nn}^{\text{SQC}}(\mathbf{e},\boldsymbol{\theta}) &\equiv K_{nn}^{\text{SQC}}(\mathbf{e}) = h\!\left(e^{(n)}-1\right) h\!\left(2-e^{(n'\neq n)}-e^{(n)}\right) \\
K_{mm}^{\text{bin}}(\mathbf{e},\boldsymbol{\theta}) &\equiv K_{mm}^{\text{bin}}(\mathbf{e}) = h\!\left(e^{(m)}-1\right) h\!\left(1-e^{(m'\neq m)}\right)
\end{aligned} \qquad (74)$$

Figure 5(a) shows the triangle window functions $\{K_{nn}^{\text{SQC}}(\mathbf{e})\}$ on the action space of TSS.

**IV-B.  Triangle Window Functions on Constraint Coordinate-Momentum Phase Space of TSS**

Using Eq. (3) and Eq. (74) to reformulate Eq. (68), for TSS, we obtain

$$p_{n\to m}^{\text{SQC}}(t) = \int \frac{2 \mathrm{d}\mathbf{e}_0 \mathrm{d}\boldsymbol{\theta}_0}{(2\pi)^2} h\!\left(e_0^{(n)}-1\right) h\!\left(2-e_0^{(n'\neq n)}-e_0^{(n)}\right) \\ \times h\!\left(e_t^{(m)}-1\right) h\!\left(1-e_t^{(m'\neq m)}\right) \qquad (75)$$

The next crucial step is to transform Eq. (75) to an equivalent formulation on the single $\mathrm{U}(F)/\mathrm{U}(F-1)$ CPS (i.e., Eq. (6)), which eliminates the product of the Heaviside step functions $h\!\left(e_t^{(m)}-1\right) h\!\left(1-e_t^{(m'\neq m)}\right)$.

The EOMs of $(\mathbf{x}_t, \mathbf{p}_t)$ or $(\mathbf{e}_t, \boldsymbol{\theta}_t)$ for TSS preserve the quantity $e_t^{(1)} + e_t^{(2)}$. That is, the phase point at any time $t$ lies on the *iso-action* line $e^{(1)} + e^{(2)} = \lambda$, where $\lambda$ is determined by the initial sum of actions.  It hints that the integral over the plane

$\{e^{(1)}, e^{(2)}\}$ can be reformulated on the $U(F)/U(F-1)$ CPS for TSS, i.e.,

$$\mathcal{S}(\mathbf{x},\mathbf{p};\gamma) = \delta\left(\sum_{n=1}^{2}\frac{1}{2}\left(\left(x^{(n)}\right)^2 + \left(p^{(n)}\right)^2\right) - (1+F\gamma)\right) = \delta\left(e^{(1)} + e^{(2)} - (1+F\gamma)\right) . \quad (76)$$

The integration on the action space can be expressed as

$$\begin{aligned}
&\int_0^{+\infty} d\bar{e}^{(1)} \int_0^{+\infty} d\bar{e}^{(2)} f\left(\bar{e}^{(1)}, \bar{e}^{(2)}; \boldsymbol{\theta}\right) \\
&= \int_0^{+\infty} d\bar{\lambda} \int_0^{+\infty} d\bar{e}^{(1)} \int_0^{+\infty} d\bar{e}^{(2)} f\left(\bar{e}^{(1)}, \bar{e}^{(2)}; \boldsymbol{\theta}\right) \delta\left(\bar{e}^{(1)} + \bar{e}^{(2)} - \bar{\lambda}\right) \\
&\underset{\substack{\bar{\lambda}=(1+F\gamma)\lambda \\ \bar{e}^{(1)}=\lambda e^{(1)} \\ \bar{e}^{(2)}=\lambda e^{(2)}}}{=} (1+F\gamma) \int_0^{+\infty} \lambda d\lambda \int_0^{+\infty} de^{(1)} \int_0^{+\infty} de^{(2)} f\left(\lambda e^{(1)}, \lambda e^{(2)}; \boldsymbol{\theta}\right) \delta\left(e^{(1)} + e^{(2)} - (1+F\gamma)\right)
\end{aligned} \quad (77)$$

It is straightforward to show that the transformation $\{\bar{e}_0^{(1)}, \bar{e}_0^{(2)}\} \mapsto \{\lambda e_0^{(1)}, \lambda e_0^{(2)}\}$ leads to $\{\bar{e}_t^{(1)}, \bar{e}_t^{(2)}\} \mapsto \{\lambda e_t^{(1)}, \lambda e_t^{(2)}\}$ at any time $t$. Applying Eq. (77) to Eq. (75) yields

$$\begin{aligned}
&p_{n\to m}^{\text{SQC}}(t) \\
&= \int \frac{2d\mathbf{e}_0 d\boldsymbol{\theta}_0}{(2\pi)^2} h\left(e_0^{(n)} - 1\right) h\left(2 - e_0^{(n'\neq n)} - e_0^{(n)}\right) h\left(e_t^{(m)} - 1\right) h\left(1 - e_t^{(m'\neq m)}\right) \\
&= (1+F\gamma) \int \frac{2d\mathbf{e}_0 d\boldsymbol{\theta}_0}{(2\pi)^2} \delta\left(e_0^{(1)} + e_0^{(2)} - (1+F\gamma)\right) \times \\
&\quad \int_0^{+\infty} \lambda d\lambda h\left(\lambda - \left(e_0^{(n)}\right)^{-1}\right) h\left(\frac{2}{e_0^{(n'\neq n)} + e_0^{(n)}} - \lambda\right) \\
&\quad h\left(\lambda - \left(e_t^{(m)}\right)^{-1}\right) h\left(\left(e_t^{(m'\neq m)}\right)^{-1} - \lambda\right) \\
&= (1+F\gamma) \int \frac{2d\mathbf{e}_0 d\boldsymbol{\theta}_0}{(2\pi)^2} \delta\left(e_0^{(1)} + e_0^{(2)} - (1+F\gamma)\right) \left[\int_{h_1}^{h_2} \lambda d\lambda\right]_+
\end{aligned} \quad (78)$$

where

$$\begin{aligned}
h_1 &= \max\left\{\left(e_0^{(n)}\right)^{-1}, \left(e_t^{(m)}\right)^{-1}\right\}, \\
h_2 &= \min\left\{\frac{2}{e_0^{(n'\neq n)} + e_0^{(n)}}, \left(e_t^{(m'\neq m)}\right)^{-1}\right\}
\end{aligned} \quad (79)$$

and $[a]_+ = \begin{cases} a, & \text{if } a > 0, \\ 0, & \text{otherwise} \end{cases}$.

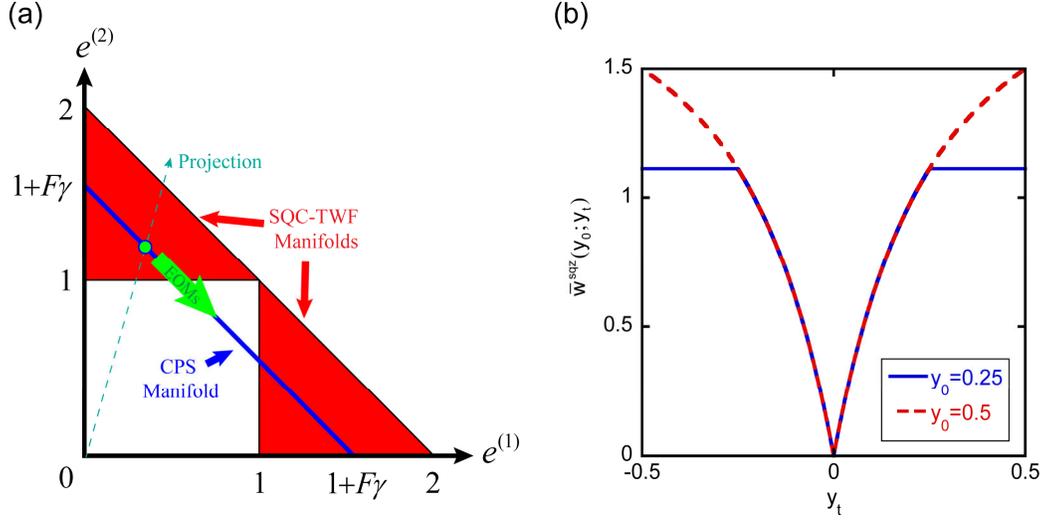

**Figure 5.** Panel (a): Illustration of the reformulation procedure of the SQC triangle window functions (TWFs) of TSS. The red triangles represent the SQC-TWF manifolds on the action space. One phase point on the SQC-TWF manifold, represented as the green circle, only moves on the CPS manifold denoted as the blue line. Thus, the integration of phase space functions on the SQC-TWF manifolds can be decomposed into integrations on different CPSs with different values of parameter $\gamma$. The projection of the integrations on all related CPSs onto the single CPS yields the squeezed SQC-TWF approach. Panel (b): Visualization of the generalized weight function $\bar{w}^{\mathrm{sqz}}(y_0; y_t)$. Red dashed line: the case where $y_0 = 0.5$. Blue solid line: the case where $y_0 = 0.25$.

Since all the Heaviside step functions only restrict the integration area of $\lambda$ in the last line of Eq. (78), we are able to reduce the dimension of the integral. The initial condition at time $0$ and the condition at time $t$ read:

At time $0$: The $n$-th state is occupied. When the value of Eq. (78) is larger than zero, we have $e_0^{(n)} > (1+F\gamma)/2$, $e_0^{(n'\neq n)} < (1+F\gamma)/2$, and $2/(e_0^{(n'\neq n)} + e_0^{(n)}) = 2/(1+F\gamma)$.

At time $t$: When the result of Eq. (78) is not zero, $\lambda - (e_t^{(m)})^{-1} > 0$ and $(e_t^{(m'\neq m)})^{-1} - \lambda > 0$, so that $e_t^{(m)} > e_t^{(m'\neq m)}$. In addition to $e_t^{(m)} + e_t^{(m'\neq m)} = 1+F\gamma$, we obtain $(e_t^{(m)})^{-1} < 2/(1+F\gamma)$ and $(e_t^{(m'\neq m)})^{-1} > 2/(1+F\gamma)$.

By employing the results above in Eq. (79), we obtain a simplified version of Eq. (78),

$$p_{n\to m}^{\text{SQC}}(t) = (1+F\gamma)\int \frac{2\mathrm{d}\mathbf{e}_0 \mathrm{d}\boldsymbol{\theta}_0}{(2\pi)^2}\delta\left(e_0^{(1)} + e_0^{(2)} - (1+F\gamma)\right)\left[\int_{h_1}^{h_2}\lambda\mathrm{d}\lambda\right]_+$$

$$= (1+F\gamma)\int \frac{2\mathrm{d}\mathbf{e}_0 \mathrm{d}\boldsymbol{\theta}_0}{(2\pi)^2}\delta\left(e_0^{(1)} + e_0^{(2)} - (1+F\gamma)\right)\times$$

$$h\left(e_0^{(n)} - \frac{1+F\gamma}{2}\right)h\left(e_t^{(m)} - \frac{1+F\gamma}{2}\right)\left(\frac{2}{(1+F\gamma)^2} - \frac{1}{2\min\{e_0^{(n)},e_t^{(m)}\}^2}\right) \quad (80)$$

By noticing

$$h\left(e_0^{(n)} - \frac{1+F\gamma}{2}\right)h\left(e_t^{(m)} - \frac{1+F\gamma}{2}\right)\left(\frac{2}{(1+F\gamma)^2} - \frac{1}{2\min\{e_0^{(n)},e_t^{(m)}\}^2}\right)$$

$$\equiv h\left(e_0^{(n)} - \frac{1+F\gamma}{2}\right)h\left(e_t^{(m)} - \frac{1+F\gamma}{2}\right)\frac{2 - \left[2\min\{|y_0|+1/2,|y_t|+1/2\}^2\right]^{-1}}{(1+F\gamma)^2}, \quad (81)$$

we express Eq. (80) with the notations defined in Eqs. (30)-(33), and obtain the following mapping formalism,

$$\mathrm{Tr}_e\left[|n\rangle\langle n|e^{i\hat{H}t}|m\rangle\langle m|e^{-i\hat{H}t}\right]$$
$$\mapsto \left(\bar{C}_{nn,mm}^{\text{sqz}}(t)\right)^{-1}\int_{\mathcal{S}(\mathbf{x}_0,\mathbf{p}_0;\gamma)} F\mathrm{d}\mathbf{x}_0\mathrm{d}\mathbf{p}_0\bar{w}^{\text{sqz}}(\mathbf{x}_0,\mathbf{p}_0;\mathbf{x}_t,\mathbf{p}_t)K_{nn}^{\text{sqz}}(\mathbf{x}_0,\mathbf{p}_0)K_{mm}^{\text{sqz}}(\mathbf{x}_t,\mathbf{p}_t), \quad (82)$$

where the generalized weight function is

$$\bar{w}^{\text{sqz}}(\mathbf{x}_0,\mathbf{p}_0;\mathbf{x}_t,\mathbf{p}_t) = \bar{w}^{\text{sqz}}(y_0(\mathbf{x}_0,\mathbf{p}_0);y_t(\mathbf{x}_t,\mathbf{p}_t))$$
$$= 2 - \frac{1}{2\min\{|y_0(\mathbf{x}_0,\mathbf{p}_0)|+1/2,|y_t(\mathbf{x}_t,\mathbf{p}_t)|+1/2\}^2}, \quad (83)$$

which depends on the values of mapping variables at both time $0$ and time $t$. The phase space functions of Eq. (82) read

$$K_{nn}^{sqz}(\mathbf{x},\mathbf{p}) = h\left(\frac{1}{2}\left(\left(x^{(n)}\right)^2 + \left(p^{(n)}\right)^2\right) - \frac{1+F\gamma}{2}\right) . \tag{84}$$

The time-dependent normalization factor is

$$\bar{C}_{nn,mm}^{sqz}(t) = \sum_{k=1}^{2} \int_{S(\mathbf{x}_0,\mathbf{p}_0;\gamma)} F d\mathbf{x}_0 d\mathbf{p}_0 \bar{w}^{sqz}(\mathbf{x}_0,\mathbf{p}_0;\mathbf{x}_t,\mathbf{p}_t) K_{nn}^{sqz}(\mathbf{x}_0,\mathbf{p}_0) K_{kk}^{sqz}(\mathbf{x}_t,\mathbf{p}_t). \tag{85}$$

We denote Eqs. (82)-(85) as the squeezed (sqz) SQC-TWF approach. The sqz approach is equivalent to the SQC-TWF approach for the pure electronic population dynamics. It is easy to verify that $\bar{w}^{sqz}(\mathbf{x}_0,\mathbf{p}_0;\mathbf{x}_t,\mathbf{p}_t)$ of the sqz approach, Eq. (83), is a case of the generalized weight function defined in Eq. (33) for the novel class of phase space representations of TSS.

**IV-C. Proof that the TWF Representation of the SQC Approach leads to the Exact Population Dynamics of TSS**

Using the generalized weight function $\bar{w}^{sqz}(\mathbf{x}_0,\mathbf{p}_0;\mathbf{x}_t,\mathbf{p}_t)$ defined by Eq. (83), we can directly employ the techniques developed in Section II and Section III to prove that the expression of Eq. (82) yields the exact population dynamics of TSS. Substitution of Eq. (83) into Eqs. (48) and (49) leads to

$$p_{1\to 1}^{\text{sqz}}(t) = \frac{2}{\pi}\int_{\Omega_{1\to 1}}\mathrm{d}y_0\mathrm{d}y_t\,\frac{\overline{w}^{\text{sqz}}(y_0;y_t)}{\sqrt{\mathcal{E}(y_0,y_t)}}$$

$$= \frac{2}{\pi}\int_{\Omega_{1\to 1}^{\text{rdn}}}\mathrm{d}y_0\mathrm{d}y_t\,\frac{1}{\sqrt{\mathcal{E}(y_0,y_t)}}\left[4-\frac{1}{(y_t+1/2)^2}\right]$$

$$= \frac{1}{\pi}\int_0^{\pi/2}\mathrm{d}s\int_{\xi(t)}^{\pi/2}\mathrm{d}T\sin(s)\left[4-\frac{4}{\left(\sin(s)\sin(T-\xi(t))+1\right)^2}\right] \quad (86)$$

$$= 1-\frac{\sin(2\xi(t))-2\xi\cos(2\xi(t))}{\pi\sin^2(\xi(t))}$$

and

$$p_{1\to 2}^{\text{sqz}}(t) = \frac{2}{\pi}\int_{\Omega_{1\to 2}}\mathrm{d}y_0\mathrm{d}y_t\,\frac{\overline{w}^{\text{sqz}}(y_0;y_t)}{\sqrt{\mathcal{E}(y_0,y_t)}}$$

$$= \frac{2}{\pi}\int_{\Omega_{1\to 2}^{\text{rdn}}}\mathrm{d}y_0\mathrm{d}y_t\,\frac{1}{\sqrt{\mathcal{E}(y_0,y_t)}}\left[4-\frac{1}{(-y_t+1/2)^2}\right]$$

$$= \frac{1}{\pi}\int_0^{\pi/2}\mathrm{d}s\int_0^{\xi(t)}\mathrm{d}T\sin(s)\left[4-\frac{4}{\left(-\sin(s)\sin(T-\xi(t))+1\right)^2}\right] \quad (87)$$

$$= \tan^2(\xi(t))-\frac{\sin(2\xi(t))-2\xi(t)\cos(2\xi(t))}{\pi\cos^2(\xi(t))}$$

Equations (86)-(87) suggest the equality

$$\frac{p_{1\to 1}^{\text{sqz}}(t)}{\cos^2(\xi(t))} = \frac{p_{1\to 2}^{\text{sqz}}(t)}{\sin^2(\xi(t))} = \overline{C}_{nn,mm}^{\text{sqz}}(t) \quad (88)$$

holds, which obeys the rule of Eq. (52). It then proves that the population-population correlation function of the sqz approach, as well as the SQC-TWF approach, is exact. The time-dependent normalization factor reads

$$\overline{C}_{nn,mm}^{\text{SQC}}(t) \equiv \overline{C}_{nn,mm}^{\text{sqz}}(t) = \frac{1}{\cos^2(\xi(t))} - \frac{4(1-2\xi(t)\cot(2\xi(t)))}{\pi\sin(2\xi(t))} \quad (89)$$

where $\overline{C}_{nn,mm}^{\text{SQC}}(0) \equiv \overline{C}_{nn,mm}^{\text{sqz}}(0) = 1$. For most cases, $\overline{C}_{nn,mm}^{\text{SQC}}(t) < 1$, which is shown by the blue lines in Figure 4. The fact that the normalization factor $\overline{C}_{nn,mm}^{\text{SQC}}(t) < 1$ indicates that for most of the time, we have a certain number of trajectories that are not counted in the population-population correlation function of the SQC-TWF approach.

V. **Conclusion.**

We have recently developed exact CPS formulations of the finite-state quantum system with covariant phase space functions in Refs. [10-18]. In this paper we employ the $\mathrm{U}(F)/\mathrm{U}(F-1)$ CPS[10, 12, 13] and non-covariant phase space functions to propose a novel class of isomorphic representations for the exact population dynamics (i.e., the exact population-population correlation function) of TSS. In the novel class defined by Eqs. (65)-(66), all phase space functions, time-dependent weight functions, and time-dependent normalization factors are positive semi-definite. As illustrated in the discussion for Eq. (65) in Subsection III.C, the contribution of each trajectory on CPS to the integral expression for the population-population correlation function is positive semi-definite by construction.

The projection of the SQC triangle window functions[28] onto the single CPS leads to a special case of the new class of representations of exact population dynamics of TSS. It proves that the triangle window function approach (of electronic DOFs), albeit empirically proposed by Cotton and Miller[28], is an exact model for reproducing the population dynamics for TSS. The analytical proof offers indispensable insight in understanding why the triangle window function approach[28] is useful in studying population dynamics of various nonadiabatic processes[23, 30, 35, 40-50]. It is expected that the novel class of isomorphic representations will encourage more fruitful investigations in developing new trajectory-based methodologies for nonadiabatic transition dynamics. For instance, the combination of the novel class of phase space representations with recently developed nonadiabatic field (NaF) dynamics[18] leads to a promising trajectory-based nonadiabatic dynamics approach[57].

**Acknowledgement**

We thank Baihua Wu, Youhao Shang, Haocheng Lu, and Bingqi Li from Peking University for useful discussions. We also thank Bill Miller from the University of California, Berkeley for having encouraged us to investigate the window function approach. This work was supported by the National Science Fund for Distinguished Young Scholars Grant No. 22225304. We acknowledge the High-performance Computing Platform of Peking University, Beijing PARATERA Tech Co., Ltd., and Guangzhou Supercomputer Center for providing computational resources.

**Conflict of Interest Statement**

There are no conflicts of interest to declare.

**Data Availability**

The data that support the findings of this study are available from the corresponding author upon reasonable request.

**Appendix A. Explicit Expression of the Time-dependent Parameter Set $(\xi, \Phi, \varphi, \psi)$**

The $2\times2$ Hermitian matrix $\mathbf{H}$ can be generally expressed as

$$\mathbf{H} = \begin{pmatrix} H_{11} & H_{12,R} + iH_{12,I} \\ H_{12,R} - iH_{12,I} & H_{22} \end{pmatrix} . \tag{90}$$

It is straightforward to derive the analytical expression of the four elements of $\mathbf{U}(t) = e^{-i\mathbf{H}t}$,

$$\begin{aligned}
U_{11}(t) &= e^{-\frac{i(H_{11}+H_{22})t}{2}} \left[ \cos\left(\frac{\sqrt{\Delta}t}{2}\right) + i\sin\left(\frac{\sqrt{\Delta}t}{2}\right) \frac{H_{22} - H_{11}}{\sqrt{\Delta}} \right] \\
U_{12}(t) &= 2e^{-\frac{i(H_{11}+H_{22})t}{2}} \sin\left(\frac{\sqrt{\Delta}t}{2}\right) \frac{H_{12,I} - iH_{12,R}}{\sqrt{\Delta}} \\
U_{21}(t) &= -2e^{-\frac{i(H_{11}+H_{22})t}{2}} \sin\left(\frac{\sqrt{\Delta}t}{2}\right) \frac{H_{12,I} + iH_{12,R}}{\sqrt{\Delta}} \\
U_{22}(t) &= e^{-\frac{i(H_{11}+H_{22})t}{2}} \left[ \cos\left(\frac{\sqrt{\Delta}t}{2}\right) - i\sin\left(\frac{\sqrt{\Delta}t}{2}\right) \frac{H_{22} - H_{11}}{\sqrt{\Delta}} \right]
\end{aligned} \tag{91}$$

where

$$\Delta = (H_{22} - H_{11})^2 + 4(H_{12,R}^2 + H_{12,I}^2) . \tag{92}$$

When $\Delta = 0$, $\mathbf{H} = H_{11}\mathbf{1}$, and $\mathbf{U}(t) = e^{-iH_{11}t}\mathbf{1}$. When $\Delta \neq 0$, the comparison of Eq. (91) to the parameterization formula

$$\mathbf{U}(t) = e^{-i\Phi} \begin{pmatrix} e^{i\psi} \cos\xi & e^{i\varphi} \sin\xi \\ -e^{-i\varphi} \sin\xi & e^{-i\psi} \cos\xi \end{pmatrix} \tag{93}$$

leads to

$$\Phi = \frac{H_{11} + H_{22}}{2} t , \tag{94}$$

and

$$\xi = \arcsin\left\{ 2\left|\sin\left(\frac{\sqrt{\Delta}}{2}t\right)\right| \sqrt{\frac{H_{12,R}^2 + H_{12,I}^2}{\Delta}} \right\} , \tag{95}$$

where $\xi$ can always be chosen in the region

$$\xi \in [0, \pi/2] . \tag{96}$$

Then, $\psi$ is the unique solution in $[0, 2\pi)$ to the following equation set:

$$\begin{cases} \cos\psi = \cos\left(\frac{\sqrt{\Delta}t}{2}\right)\sqrt{\dfrac{\Delta}{\Delta - 4\left(H_{12,R}^2 + H_{12,I}^2\right)\sin^2\left(\frac{\sqrt{\Delta}t}{2}\right)}} \\ \sin\psi = \sin\left(\frac{\sqrt{\Delta}t}{2}\right)\dfrac{(H_{22} - H_{11})}{\sqrt{\Delta - 4\left(H_{12,R}^2 + H_{12,I}^2\right)\sin^2\left(\frac{\sqrt{\Delta}t}{2}\right)}} \end{cases} \quad (97)$$

and $\varphi$ is the unique solution in $[0, 2\pi)$ to the following equation set when $\sin\left(\frac{\sqrt{\Delta}t}{2}\right) \neq 0$:

$$\begin{cases} \cos\varphi = \operatorname{sgn}\left(\sin\left(\frac{\sqrt{\Delta}t}{2}\right)\right)\dfrac{H_{12,I}}{\sqrt{H_{12,R}^2 + H_{12,I}^2}} \\ \sin\varphi = -\operatorname{sgn}\left(\sin\left(\frac{\sqrt{\Delta}t}{2}\right)\right)\dfrac{H_{12,R}}{\sqrt{H_{12,R}^2 + H_{12,I}^2}} \end{cases} \quad (98)$$

It is important to understand that $\Phi$ varies with time $t$, as suggested by Eq. (94). Without $e^{-i\Phi}$, the parameterization similar to Eq. (22) was explicitly used in Ref. [53] and could be traced back as early as Ref. [58] in 1897. The expression of the evolution matrix similar to Eq. (22) with $e^{-i\Phi}$ is often employed in quantum computing and information[54, 55].

**Appendix B. Transformation of Variables from $(\mathbf{x}_0, \mathbf{p}_0)$ to $(y_0, \theta_0^d)$ for Integrals over Constraint Phase Space of TSS**

Consider the bounded integrand function $\sigma\left(y_0(\mathbf{x}_0, \mathbf{p}_0); y_t(\mathbf{x}_t, \mathbf{p}_t)\right) \equiv \sigma\left(y_0; y_t(y_0, \theta_0^d)\right)$ on CPS. Using Eqs. (12)-(13) and Eq. (3), the transformation between the integral over the coordinate-momentum variables and its counterpart over the action-angle variables of TSS reads

$$\int_{S(\mathbf{x}_0,\mathbf{p}_0;\gamma)} d\mathbf{x}_0 d\mathbf{p}_0 \, \sigma\big(y_0(\mathbf{x}_0,\mathbf{p}_0); y_t(\mathbf{x}_t,\mathbf{p}_t)\big)$$

$$= \frac{1}{(2\pi)^2 (1+F\gamma)} \int_0^{+\infty} de_0^{(1)} \int_0^{+\infty} de_0^{(2)} \delta\big(e_0^{(1)} + e_0^{(2)} - (1+F\gamma)\big) \quad . \tag{99}$$

$$\times \int_0^{2\pi} d\theta_0^{(1)} \int_0^{2\pi} d\theta_0^{(2)} \sigma\big(y_0; y_t(y_0, \theta_0^d)\big)$$

By utilizing Eq. (25) and Eq. (27), and separating the integral area $\big(\theta_0^{(1)}, \theta_0^{(2)}\big) \in [0, 2\pi) \otimes [0, 2\pi)$ into two regions, one for $\theta_0^{(1)} \leq \theta_0^{(2)}$ and the other for $\theta_0^{(1)} > \theta_0^{(2)}$, we obtain

$$\int_0^{2\pi} d\theta_0^{(1)} \int_0^{2\pi} d\theta_0^{(2)} \sigma\big(y_0; y_t(y_0, \theta_0^d)\big)$$
$$= \int_0^{2\pi} d\theta_0^{(1)} \int_0^{\theta_0^{(1)}} d\theta_0^{(2)} \sigma\big(y_0; y_t(y_0, \theta_0^d)\big) + \int_0^{2\pi} d\theta_0^{(1)} \int_{\theta_0^{(1)}}^{2\pi} d\theta_0^{(2)} \sigma\big(y_0; y_t(y_0, \theta_0^d)\big)$$
$$= \int_0^{2\pi} d\theta_0^{(1)} \int_{2\pi-\theta_0^{(1)}}^{2\pi} d\theta_0^d \sigma\big(y_0; y_t(y_0, \theta_0^d)\big) + \int_0^{2\pi} d\theta_0^{(1)} \int_0^{2\pi-\theta_0^{(1)}} d\theta_0^d \sigma\big(y_0; y_t(y_0, \theta_0^d)\big) \quad . \tag{100}$$
$$= \int_0^{2\pi} d\theta_0^{(1)} \int_0^{2\pi} d\theta_0^d \sigma\big(y_0; y_t(y_0, \theta_0^d)\big)$$
$$= 2\pi \int_0^{2\pi} d\theta_0^d \sigma\big(y_0; y_t(y_0, \theta_0^d)\big)$$

Substitution of Eq. (100) and Eq. (28) into Eq. (99) produces

$$\int_{S(\mathbf{x}_0,\mathbf{p}_0;\gamma)} d\mathbf{x}_0 d\mathbf{p}_0 \, \sigma\big(y_0(\mathbf{x}_0,\mathbf{p}_0); y_t(\mathbf{x}_t,\mathbf{p}_t)\big)$$
$$= \frac{1}{(2\pi)(1+F\gamma)} \int_0^{+\infty} de_0^{(1)} \int_0^{+\infty} de_0^{(2)} \delta\big(e_0^{(1)} + e_0^{(2)} - (1+F\gamma)\big) \int_0^{2\pi} d\theta_0^d \sigma\big(y_0; y_t(y_0, \theta_0^d)\big). \tag{101}$$
$$= \frac{1}{2\pi} \int_{-1/2}^{1/2} dy_0 \int_0^{2\pi} d\theta_0^d \sigma\big(y_0; y_t(y_0, \theta_0^d)\big)$$

Equation (101) holds for general expressions of $\sigma\big(y_0(\mathbf{x}_0,\mathbf{p}_0); y_t(\mathbf{x}_t,\mathbf{p}_t)\big)$. When we use $\sigma(y_0; y_t) = F\overline{w}(y_0; y_t) K_{nn}(y_0) K_{mm}(y_t)$ in Eq. (101), it is straightforward to show that Eqs. (30)-(33) lead to Eq. (38) of the main text.

**Appendix C. Exact Population Dynamics for the Special Case of** $\sin(2\xi(t)) = 0$

From Eq. (29), when $\sin(2\xi(t)) = 0$, $|y_t| = |y_0|$. Thus, Eq. (38) leads to

$$p_{n \to m}(t) = \frac{1}{\pi} \int_{-1/2}^{1/2} dy_0 \int_0^{2\pi} d\theta_0^d f\big(|y_0|\big) K_{nn}(y_0) K_{mm}(y_t) \quad . \tag{102}$$

The circumstances where $y_t = y_0$ and $y_t = -y_0$ will separately be discussed.

When $y_t = y_0$, Eq. (102) becomes

$$p_{n\to m}(t) = 2\int_{-1/2}^{1/2} dy_0 f(|y_0|) K_{nn}(y_0) K_{mm}(y_0) \ . \tag{103}$$

Substitution of the property

$$h(y)h(-y) \equiv 0 \tag{104}$$

into Eq. (103) leads to $p_{1\to 2}(t) = p_{2\to 1}(t) = 0$, thus

$$\begin{aligned}\overline{C}_{11,11}(t) = \overline{C}_{11,22}(t) = p_{1\to 1}(t), \\ \overline{C}_{22,11}(t) = \overline{C}_{22,22}(t) = p_{2\to 2}(t)\end{aligned} \tag{105}$$

in the second and third lines of Eq. (30). We thus obtain

$$\begin{pmatrix} \left(\overline{C}_{11,11}(t)\right)^{-1} p_{1\to 1}(t) & \left(\overline{C}_{11,22}(t)\right)^{-1} p_{1\to 2}(t) \\ \left(\overline{C}_{22,11}(t)\right)^{-1} p_{2\to 1}(t) & \left(\overline{C}_{22,22}(t)\right)^{-1} p_{2\to 2}(t) \end{pmatrix} = \begin{pmatrix} 1 & 0 \\ 0 & 1 \end{pmatrix} \ . \tag{106}$$

Because $\xi \in [0, \pi/2]$, Eq. (29) suggests that $y_t = y_0$ corresponds to $\xi = 0$. Substitution of $\xi = 0$ into Eq. (22) produces

$$\begin{pmatrix} |U_{11}(t)|^2 & |U_{21}(t)|^2 \\ |U_{12}(t)|^2 & |U_{22}(t)|^2 \end{pmatrix} = \begin{pmatrix} 1 & 0 \\ 0 & 1 \end{pmatrix} \ . \tag{107}$$

Substituting Eq. (107) into the LHS of Eq. (37) and Eq. (106) into the RHS of the first line of Eq. (37), we prove that the equality of Eq. (37), the condition of the exact population dynamics, holds for the $y_t = y_0$ case.

When $y_t = -y_0$, Eq. (102) leads to

$$p_{n\to m}(t) = 2\int_{-1/2}^{1/2} dy_0 f(|y_0|) K_{nn}(y_0) K_{mm}(-y_0) \tag{108}$$

such that

$$p_{1\to 1}(t) = p_{2\to 2}(t) = 0 \ . \tag{109}$$

Substitution of Eq. (109) into the RHS of the third line of Eq. (30) leads to

$$\bar{C}_{11,11}(t) = \bar{C}_{11,22}(t) = p_{1\to 2}(t),$$
$$\bar{C}_{22,11}(t) = \bar{C}_{22,22}(t) = p_{2\to 1}(t)$$
(110)

which produces

$$\begin{pmatrix} \left(\bar{C}_{11,11}(t)\right)^{-1} p_{1\to 1}(t) & \left(\bar{C}_{11,22}(t)\right)^{-1} p_{1\to 2}(t) \\ \left(\bar{C}_{22,11}(t)\right)^{-1} p_{2\to 1}(t) & \left(\bar{C}_{22,22}(t)\right)^{-1} p_{2\to 2}(t) \end{pmatrix} = \begin{pmatrix} 0 & 1 \\ 1 & 0 \end{pmatrix}.$$
(111)

In $\xi \in [0, \pi/2]$, Eq. (29) leads to that $y_t = -y_0$ corresponds to $\xi = \pi/2$. Substitution of $\xi = \pi/2$ into Eq. (22) yields

$$\begin{pmatrix} |U_{11}(t)|^2 & |U_{21}(t)|^2 \\ |U_{12}(t)|^2 & |U_{22}(t)|^2 \end{pmatrix} = \begin{pmatrix} 0 & 1 \\ 1 & 0 \end{pmatrix}.$$
(112)

Substituting Eq. (112) into the LHS of Eq. (37) and Eq. (111) into the RHS of the first line of Eq. (37), we verify that the equality of Eq. (37), the condition of the exact population dynamics, holds for the $y_t = -y_0$ case.

As a summary, in either of the two cases, $y_t = \pm y_0$ derived from Eq. (29) when $\sin(2\xi(t)) = 0$, the mapping formalism of Eqs. (30)-(33) satisfies the equality of Eq. (37), the condition of the exact population dynamics of TSS, which is essentially the heart of the proof.

**Appendix D. Integral Identity of Equation (39) for the Non-trivial Case of $\sin(2\xi(t)) \neq 0$**

We aim to transform the integral $\frac{1}{2\pi} \int_{-1/2}^{1/2} dy_0 \int_0^{2\pi} d\theta_0^d \sigma\left(y_0; y_t\left(y_0, \theta_0^d\right)\right)$, which is the last line of Eq. (101) of Appendix B, to an integral of variables $(y_0, y_t)$. The dependence of $y_t$ upon $\theta_0^d$ is via $\cos(\varphi - \psi + \theta_0^d)$. The property of the cosine function indicates that $\cos(\varphi - \psi + \theta_0^d)$ monotonically depends on $\theta_0^d$ on at least one interval whose length is $\pi$ in the domain $\theta_0^d \in [0, 2\pi)$. Denote such an interval as $V_{\theta_0^d}$. As shown in Figure 6, it is easy to determine $V_{\theta_0^d}$ for a chosen value of $\varphi-\psi$. The value of $\varphi-\psi$ lies in region $(-2\pi, 2\pi)$.

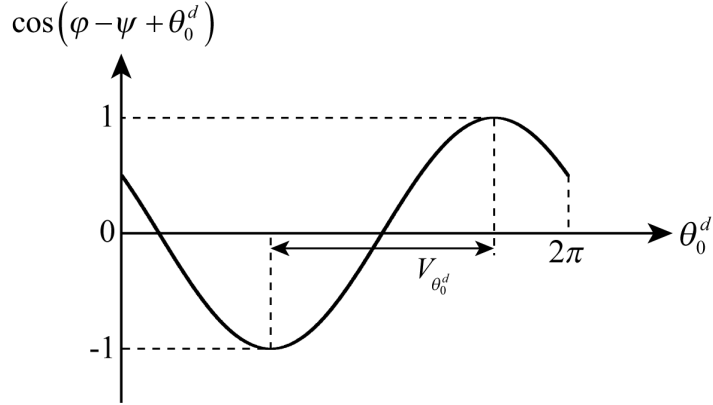

**Figure 6.** The function, $\cos(\varphi-\psi+\theta_0^d)$ for $\theta_0^d \in [0, 2\pi)$. Here, $\varphi-\psi = \pi/3$ is used for demonstration only. On the interval of $V_{\theta_0^d}$, $\cos(\varphi-\psi+\theta_0^d)$ is monotonic. The length of interval $V_{\theta_0^d}$ is $\pi$. It is straightforward to determine interval $V_{\theta_0^d}$ for any value of $\varphi-\psi$.

Because of Eq. (29) and the symmetry of the cosine function, it is straightforward to show

$$\frac{1}{2\pi}\int_{-1/2}^{1/2} dy_0 \int_0^{2\pi} d\theta_0^d \sigma\left(y_0; y_t(y_0, \theta_0^d)\right) = \frac{1}{\pi}\int_{-1/2}^{1/2} dy_0 \int_{V_{\theta_0^d}} d\theta_0^d \sigma\left(y_0; y_t(y_0, \theta_0^d)\right). \quad (113)$$

Since $\sigma\left(y_0; y_t(y_0, \theta_0^d)\right)$ is a bounded function, the integration domain of the RHS of Eq. (113) is set to be $(y_0, \theta_0^d) \in (-1/2, 1/2) \otimes V_{\theta_0^d}$, where $V_{\theta_0^d}$ is an open interval whose length is $\pi$. Given that $\sin(2\xi) \neq 0$ and the value of $y_0$ is fixed, Eq. (29) suggests that $y_t$ monotonically depends on $\theta_0^d$ for $\theta_0^d \in V_{\theta_0^d}$. Thus, there exists the one-to-one correspondence between $(y_0, \theta_0^d)$ and $(y_0, y_t)$ for $(y_0, \theta_0^d) \in (-1/2, 1/2) \otimes V_{\theta_0^d}$. As shown in Figure 6, in domain $V_{\theta_0^d}$, $\cos(\varphi - \psi + \theta_0^d) \in (-1, 1)$. Thus, for a chosen value of $y_0$, the interval of $\theta_0^d \in V_{\theta_0^d}$ is mapped to the corresponding interval

$$y_t \in \left( y_0 \cos(2\xi) - \sqrt{1/4 - (y_0)^2} |\sin(2\xi)|, y_0 \cos(2\xi) + \sqrt{1/4 - (y_0)^2} |\sin(2\xi)| \right). \quad (114)$$

Subtracting $y_0 \cos(2\xi)$ in both sides of Eq. (114) produces

$$y_t - y_0 \cos(2\xi) \in \left( -\sqrt{1/4 - (y_0)^2} |\sin(2\xi)|, \sqrt{1/4 - (y_0)^2} |\sin(2\xi)| \right). \quad (115)$$

Squaring both sides of Eq. (115) then defines the integration region of $(y_0, y_t)$

$$\Omega := \{(y_0, y_t) \mid \mathcal{E}(y_0, y_t) > 0\}, \quad (116)$$

where

$$\mathcal{E}(y_0, y_t) = \frac{\sin^2(2\xi)}{4} - (y_0)^2 - (y_t)^2 + 2 y_0 y_t \cos(2\xi). \quad (117)$$

It is easy to obtain the following Jacobian determinant,

$$\frac{\partial(y_0, y_t)}{\partial(y_0, \theta_0^d)} = \frac{\partial y_0}{\partial y_0} \frac{\partial y_t}{\partial \theta_0^d} - \frac{\partial y_0}{\partial \theta_0^d} \frac{\partial y_t}{\partial y_0}$$

$$= -\sqrt{\frac{1}{4} - (y_0)^2} \sin(2\xi) \sin(\varphi - \psi + \theta_0^d) \quad (118)$$

Substituting Eq. (29) into Eq. (118), we obtain

$$\left| \frac{\partial(y_0, \theta_0^d)}{\partial(y_0, y_t)} \right| = \left| \sqrt{\frac{1}{4} - (y_0)^2} \sin(2\xi) \sqrt{1 - \left( \frac{y_t - y_0 \cos(2\xi)}{\sqrt{\frac{1}{4} - (y_0)^2} \sin(2\xi)} \right)^2} \right|^{-1} = \frac{1}{\sqrt{\mathcal{E}(y_0, y_t)}}. \quad (119)$$

Substitution of Eqs. (116)-(119) into Eq. (113) leads to the integral identity of TSS

$$\frac{1}{2\pi} \int_{-1/2}^{1/2} dy_0 \int_0^{2\pi} d\theta_0^d \sigma(y_0; y_t) = \frac{1}{\pi} \int_\Omega dy_0 dy_t \frac{\sigma(y_0; y_t)}{\sqrt{\mathcal{E}(y_0, y_t)}}, \quad (120)$$

where $\Omega$ is the ellipse defined in Eqs. (116)-(117). Similar to Appendix B, we let

$\sigma(y_0; y_t) = F\overline{w}(y_0; y_t) K_{nn}(y_0) K_{mm}(y_t)$ so that Eq. (120) becomes Eq. (39) of the main text.

**Appendix E.  Equality of Elements of $\{\overline{C}_{nn,mm}(t)\}$  for All  $n$  and  $m$**

We first consider the case of $\sin(2\xi(t)) = 0$. When $y_t = y_0$, from Eq. (103) we obtain

$$\begin{cases} p_{1\to 1}(t) = 2\int_{-1/2}^{1/2} dy_0 f(|y_0|) K_{11}(y_0) K_{11}(y_0) \\ \qquad\qquad = 2\int_0^{1/2} dy_0 f(y_0), \\ p_{2\to 2}(t) = 2\int_{-1/2}^{1/2} dy_0 f(|y_0|) K_{22}(y_0) K_{22}(y_0) \\ \qquad\qquad = 2\int_{-1/2}^0 dy_0 f(-y_0) = p_{1\to 1}(t) \end{cases} \qquad (121)$$

Equation (121) and Eq. (105) of Appendix C lead to

$\overline{C}_{11,11}(t) = \overline{C}_{11,22}(t) = \overline{C}_{22,11}(t) = \overline{C}_{22,22}(t)$.

When $y_t = -y_0$, from Eq. (108) we achieve

$$\begin{cases} p_{1\to 2}(t) = 2\int_{-1/2}^{1/2} dy_0 f(|y_0|) K_{11}(y_0) K_{22}(-y_0) \\ \qquad\qquad = 2\int_0^{1/2} dy_0 f(y_0), \\ p_{2\to 1}(t) = 2\int_{-1/2}^{1/2} dy_0 f(|y_0|) K_{22}(y_0) K_{11}(-y_0) \\ \qquad\qquad = 2\int_{-1/2}^0 dy_0 f(-y_0) = p_{1\to 2}(t) \end{cases} \qquad (122)$$

Equation (122) and Eq. (110) yield $\overline{C}_{11,11}(t) = \overline{C}_{11,22}(t) = \overline{C}_{22,11}(t) = \overline{C}_{22,22}(t)$.

We then focus on the non-trivial case of $\sin(2\xi(t)) \neq 0$. From Eqs. (33), (40) and (41), we obtain $\overline{w}(y_0; y_t) = \overline{w}(-y_0; -y_t)$, $\mathcal{E}(y_0, y_t) = \mathcal{E}(-y_0, -y_t)$, and that $\Omega$ exhibits the central symmetry with respect to the point, $(y_0, y_t) = (0,0)$. Thus, we achieve

$$\begin{cases} p_{1\to1}(t) = \dfrac{2}{\pi}\int_\Omega \mathrm{d}y_0 \mathrm{d}y_t \, \dfrac{\overline{w}(y_0;y_t)h(y_0)h(y_t)}{\sqrt{\mathcal{E}(y_0,y_t)}} \\ \qquad\quad = \dfrac{2}{\pi}\int_\Omega \mathrm{d}y_0 \mathrm{d}y_t \, \dfrac{\overline{w}(y_0;y_t)h(-y_0)h(-y_t)}{\sqrt{\mathcal{E}(y_0,y_t)}} = p_{2\to2}(t), \\ p_{1\to2}(t) = \dfrac{2}{\pi}\int_\Omega \mathrm{d}y_0 \mathrm{d}y_t \, \dfrac{\overline{w}(y_0;y_t)h(y_0)h(-y_t)}{\sqrt{\mathcal{E}(y_0,y_t)}} \\ \qquad\quad = \dfrac{2}{\pi}\int_\Omega \mathrm{d}y_0 \mathrm{d}y_t \, \dfrac{\overline{w}(y_0;y_t)h(-y_0)h(y_t)}{\sqrt{\mathcal{E}(y_0,y_t)}} = p_{2\to1}(t) \end{cases} \quad (123)$$

Equation (123) in addition to

$$\begin{cases} \overline{C}_{11,11}(t) = \overline{C}_{11,22}(t) = p_{1\to1}(t) + p_{1\to2}(t), \\ \overline{C}_{22,11}(t) = \overline{C}_{22,22}(t) = p_{2\to1}(t) + p_{2\to2}(t) \end{cases} \quad (124)$$

yields $\overline{C}_{11,11}(t) = \overline{C}_{11,22}(t) = \overline{C}_{22,11}(t) = \overline{C}_{22,22}(t)$ for the case of $\sin(2\xi(t)) \neq 0$. In conclusion, for all $n$ and $m$, the elements of $\{\overline{C}_{nn,mm}(t)\}$ are the same for the mapping formalism of Eqs. (30)-(33) of TSS.

**Appendix F. Property and Condition for Boundedness of $f(y)$ Defined by Equation (64)**

In Subsection III.C of the main text, we show that Eq. (37), the condition for the exact population dynamics of TSS with the mapping formalism of Eqs. (30)-(33), which is equivalent to the axial symmetry of Eq. (53), leads to

$$f(y) = \int_0^{2y} \frac{\mathrm{d}z}{\sqrt{(2y)^2 - z^2}} \frac{\mathrm{d}B(z)}{\mathrm{d}z} \quad (125)$$

in the formula of the generalized weight function of Eq. (33), with

$$B(z) = z\sqrt{1-z^2}\, \frac{\mathrm{d}}{\mathrm{d}z}\left(z^2 \Xi(\arcsin z)\right) \quad (126)$$

where the axial symmetry of $\Xi\left(\dfrac{\pi}{2} - \xi\right) = \Xi(\xi)$ holds. Substitution of Eq. (126) into Eq. (125) leads to

$$f(y) = \int_0^{2y} \frac{dz}{\sqrt{(2y)^2 - z^2}} \frac{d}{dz}\left(z\sqrt{1-z^2} \frac{d}{dz}\left(z^2 \Xi(\arcsin z)\right)\right)$$
$$= \int_0^{2y} \frac{z\,dz}{\sqrt{(2y)^2 - z^2}} \chi(z) \tag{127}$$

with

$$\chi(z) = \frac{4 - 6z^2}{\sqrt{1-z^2}} \Xi(\arcsin z) + \left(5z\sqrt{1-z^2} - \frac{z^3}{\sqrt{1-z^2}}\right) \frac{d\Xi(\arcsin z)}{dz}$$
$$+ z^2 \sqrt{1-z^2} \frac{d^2 \Xi(\arcsin z)}{dz^2} \tag{128}$$

Similar to Subsection III.C of the main text, we define $z = \sin\xi$. It is straightforward to show

$$\begin{cases} \dfrac{d\Xi(\arcsin z)}{dz} = \dfrac{1}{\cos\xi} \dfrac{d\Xi(\xi)}{d\xi} = \dfrac{1}{\sqrt{1-z^2}} \dfrac{d\Xi(\xi)}{d\xi}, \\ \dfrac{d^2 \Xi(\arcsin z)}{dz^2} = \dfrac{1}{\cos\xi} \dfrac{d}{d\xi}\left(\dfrac{1}{\cos\xi} \dfrac{d\Xi(\xi)}{d\xi}\right) \\ \qquad = \dfrac{1}{\sqrt{1-z^2}}\left(\dfrac{z}{1-z^2} \dfrac{d\Xi(\xi)}{d\xi} + \dfrac{1}{\sqrt{1-z^2}} \dfrac{d^2 \Xi(\xi)}{d\xi^2}\right) \end{cases} \tag{129}$$

Substitution of Eq. (129) into Eq. (127) leads to

$$f(y)$$
$$= \int_0^{2y} \frac{z\,dz}{\sqrt{(2y)^2 - z^2}} \chi(z)$$
$$= \int_0^{2y} \frac{z\,dz}{\sqrt{(2y)^2 - z^2}} \left[\frac{4-6z^2}{\sqrt{1-z^2}} \Xi(\arcsin z) + 5z \left.\frac{d\Xi(\xi)}{d\xi}\right|_{z=\sin\xi} + \frac{z^2}{\sqrt{1-z^2}} \left.\frac{d^2\Xi(\xi)}{d\xi^2}\right|_{z=\sin\xi}\right] \tag{130}$$

Since we demand that $\Xi(\xi)$ is a bounded smooth function, any order of derivative of $\Xi(\xi)$ is bounded for $\xi \in [0, \pi/2]$. Thus, as long as $0 \le z < 1$ such that $1/\sqrt{1-z^2} < +\infty$, we can always set an upper bound for $\chi(z)$ of Eq. (128), or equivalently, for the term in $[\cdots]$ of Eq. (130), i.e.,

$$\left| \frac{4-6z^2}{\sqrt{1-z^2}} \Xi(\arcsin z) + 5z \frac{d\Xi(\xi)}{d\xi}\bigg|_{z=\sin\xi} \right| + \left| \frac{z^2}{\sqrt{1-z^2}} \frac{d^2\Xi(\xi)}{d\xi^2}\bigg|_{z=\sin\xi} \right| < M_{max} \qquad (131)$$

where $M_{max}$ is a large finite number. Thus, when $z < 1$ in the integral of Eq. (130), which means that for $y < 1/2$,

$$|f(y)| < \int_0^{2y} \frac{z dz}{\sqrt{(2y)^2 - z^2}} M_{max} = 2M_{max} y. \qquad (132)$$

Straightforwardly, $\lim_{y \to 0+} f(y) = 0$, and $f(y)$ is bounded for $y < 1/2$.

For $y = 1/2$,

$$f(1/2) = \int_0^1 z dz \left[ \frac{4-6z^2}{1-z^2} \Xi(\arcsin z) + \frac{5z}{\sqrt{1-z^2}} \frac{d\Xi(\xi)}{d\xi}\bigg|_{z=\sin\xi} + \frac{z^2}{1-z^2} \frac{d^2\Xi(\xi)}{d\xi^2}\bigg|_{z=\sin\xi} \right]. \qquad (133)$$

The change of variable $z = \sin\xi$ in Eq. (133) produces

$$f(1/2) = \int_0^{\pi/2} \sin\xi d\xi \left[ \frac{4\cos^2\xi - 2\sin^2\xi}{\cos\xi} \Xi(\xi) + 5\sin\xi \frac{d\Xi(\xi)}{d\xi} + \frac{\sin^2\xi}{\cos\xi} \frac{d^2\Xi(\xi)}{d\xi^2} \right]. \qquad (134)$$

Denote $\tilde{\xi} = \frac{\pi}{2} - \xi$. By using $\Xi\left(\frac{\pi}{2} - \xi\right) = \Xi(\xi)$, we have $\Xi'(\xi)\big|_{\xi=\pi/2-\tilde{\xi}} = -\Xi'(\tilde{\xi})$ and $\Xi''(\xi)\big|_{\xi=\pi/2-\tilde{\xi}} = \Xi''(\tilde{\xi})$. Equation (134) becomes

$$f(1/2) = \int_0^{\pi/2} \cos\tilde{\xi} d\tilde{\xi} \left[ \frac{4\sin^2\tilde{\xi} - 2\cos^2\tilde{\xi}}{\sin\tilde{\xi}} \Xi(\tilde{\xi}) - 5\Xi'(\tilde{\xi})\cos\tilde{\xi} + \frac{\cos^2\tilde{\xi}}{\sin\tilde{\xi}} \Xi''(\tilde{\xi}) \right]. \qquad (135)$$

The integrand function of Eq. (135), denoted as

$$A(\tilde{\xi}) = \cos\tilde{\xi} \left[ \frac{4\sin^2\tilde{\xi} - 2\cos^2\tilde{\xi}}{\sin\tilde{\xi}} \Xi(\tilde{\xi}) - 5\Xi'(\tilde{\xi})\cos\tilde{\xi} + \frac{\cos^2\tilde{\xi}}{\sin\tilde{\xi}} \Xi''(\tilde{\xi}) \right], \qquad (136)$$

is bounded for $\tilde{\xi} \in (0, \pi/2]$. The convergence of $A(\tilde{\xi})$ at the point, $\tilde{\xi} = 0$, must be carefully examined. Since $\Xi(\tilde{\xi})$ is a bounded smooth function, we expand $A(\tilde{\xi})$ around $\tilde{\xi} = 0$:

$$A(\tilde{\xi})$$
$$= \cos\tilde{\xi}\left[\frac{4\sin^2\tilde{\xi}-2\cos^2\tilde{\xi}}{\sin\tilde{\xi}}\Xi(\tilde{\xi})-5\Xi'(\tilde{\xi})\cos\tilde{\xi}+\frac{\cos^2\tilde{\xi}}{\sin\tilde{\xi}}\Xi''(\tilde{\xi})\right] \qquad (137)$$
$$=\frac{(\Xi''(0)-2\Xi(0))}{\tilde{\xi}}+(\Xi'''(0)-7\Xi'(0))+\frac{1}{6}(40\Xi(0)-44\Xi''(0)+3\Xi^{(4)}(0))\tilde{\xi}\cdots$$

When $\Xi''(0) \ne 2\Xi(0)$, for a very small $\epsilon$, the integral $\int_0^\epsilon A(\tilde{\xi})\mathrm{d}\tilde{\xi}$ is controlled by the $1/\tilde{\xi}$ term in Eq. (137), i.e.,

$$\int_0^\epsilon A(\tilde{\xi})\mathrm{d}\tilde{\xi} \sim \int_0^\epsilon \frac{(\Xi''(0)-2\Xi(0))}{\tilde{\xi}}\mathrm{d}\tilde{\xi} = (\Xi''(0)-2\Xi(0))\ln\tilde{\xi}\Big|_0^\epsilon , \qquad (138)$$

which does not converge. Thus, when $\Xi''(0) \ne 2\Xi(0)$, $f(1/2)$ is unbounded. Only when $\Xi''(0) = 2\Xi(0)$, $A(\tilde{\xi})$ is bounded for $\tilde{\xi} \in [0, \pi/2]$, and $f(y)$ is bounded for $y \in [0, 1/2]$.

**Appendix G. Expressions of $f(y)$ and $\Xi(\xi)$ in Figure 4**

In Figure 4(a), the example function of $f(y)$ reads

$$\begin{aligned}f(y) &= \frac{23}{4}y - 36y^2 + 70y^3 + 240y^4 - 420y^5 \\ &\quad + \left(\frac{9}{8}-\frac{9}{2}y^2 - 210y^4 + 840y^6\right)\mathrm{arctanh}(2y)\end{aligned}, \qquad (139)$$

and the corresponding time-dependent normalization factor is

$$\bar{C}_{nn,mm}(t) \equiv \Xi(\xi) = \frac{3-\cos^4\xi - \sin^4\xi - 4\sin(2\xi)/\pi}{2} . \qquad (140)$$

In Figure 4(b), the example function of $f(y)$ reads

$$\begin{aligned}f(y) &= 2y + 18y^2 + 128y^3 - 120y^4 + (3-36y^2)\mathrm{arctanh}(2y) \\ &\quad + (2-64y^2)\mathrm{E}(4y^2) - (2-32y^2)\mathrm{K}(4y^2)\end{aligned}, \qquad (141)$$

and the corresponding time-dependent normalization factor is

$$\bar{C}_{nn,mm}(t) \equiv \Xi(\xi) = 3 - 2\cos\xi - 2\sin\xi + \sin(2\xi)/\pi . \qquad (142)$$

In Eq. (141), $K(4y^2)$ is the complete elliptic integral of the first kind,

$$K(4y^2) = \int_0^{\pi/2} \frac{d\theta}{\sqrt{1-4y^2 \sin^2 \theta}}, \tag{143}$$

and $E(4y^2)$ is the complete elliptic integral of the second kind,

$$E(4y^2) = \int_0^{\pi/2} \sqrt{1-4y^2 \sin^2 \theta}\, d\theta. \tag{144}$$

Although the expressions of these example functions of $f(y)$ seem complicated, they are constructed from the simple generating rule of Eq. (36).

**Appendix H. Numerical Examples of TSS**

We test the squeezed (sqz) SQC-TWF approach and the example Case 1 of the novel representations of TSS. The Hamiltonian operator for the TSS model reads

$$\hat{H} = H_{11}|1\rangle\langle 1| + H_{22}|2\rangle\langle 2| + H_{12}(|1\rangle\langle 2| + |2\rangle\langle 1|). \tag{145}$$

The diagonal terms in Eq. (145) are

$$H_{11} = 10, H_{22} = 2 \tag{146}$$

and the off-diagonal term is

$$H_{12} = \lambda. \tag{147}$$

The value of the parameter $\lambda$ is set to be 0.02, 0.2, 2 and 20 from weak coupling region to strong coupling region. The parameters of the model come from a TSS reduced from the three-state model of Ref. [10].

The initial electronic density matrix is chosen to be $|1\rangle\langle 1|$, whose phase space counterpart function reads $h\left(\frac{1}{2}(x^{(1)})^2 + \frac{1}{2}(p^{(1)})^2 - \frac{1}{2}(x^{(2)})^2 - \frac{1}{2}(p^{(2)})^2\right)$. The initial conditions of the squeezed SQC-TWF approach and the example Case 1 of the novel representations are generated by the uniform sampling on the sub-manifold

$$h\left(\frac{1}{2}\left(x^{(1)}\right)^2 + \frac{1}{2}\left(p^{(1)}\right)^2 - \frac{1}{2}\left(x^{(2)}\right)^2 - \frac{1}{2}\left(p^{(2)}\right)^2\right) \times \mathcal{S}(\mathbf{x},\mathbf{p};\gamma). \tag{148}$$

The evolution of phase space variables $(\mathbf{x}_t, \mathbf{p}_t)$ corresponds to the TDSE of Eq. (20). For the two representations used in the tests, $2 \times 10^7$ trajectories are used for convergence. Figure 7 presents the population difference between two states for the TSS model as a function of time. All of the tests verify that both electronic representations lead to the exact population dynamics in various coupling regimes of the TSS model.

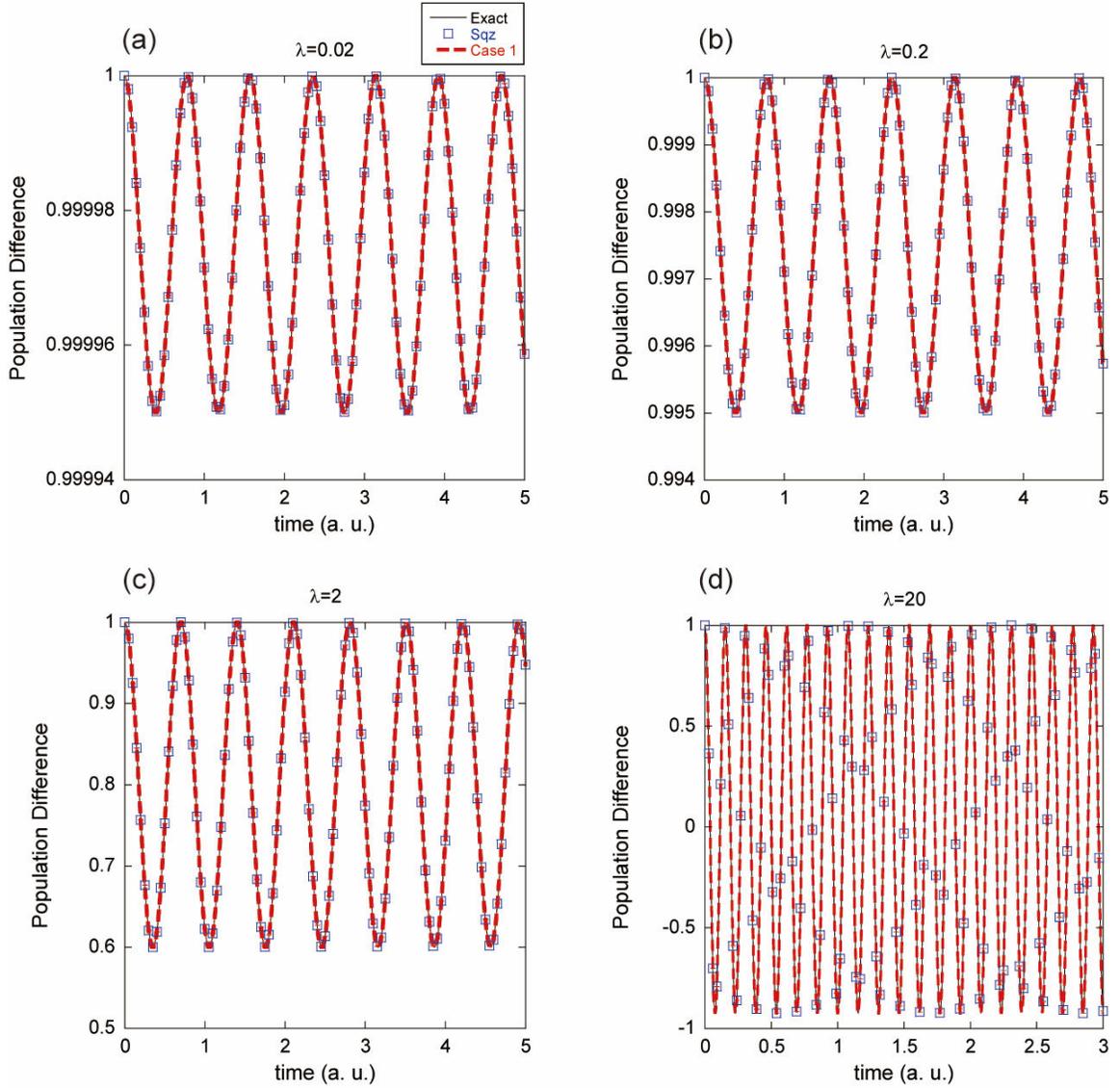

**Figure 7.** Results of population difference $P_1(t) - P_2(t)$ between the two states of the TSS model, where the initial state is $|1\rangle$. Panel (a) reports results of the TSS model with parameter $\lambda = 0.02$. Black solid line: exact results. Blue Squares: results of squeezed (sqz) SQC-TWF. Red dashed line: results of Case 1. Panels (b), (c), and (d) are similar to Panel (a) but for $\lambda = 0.2$, $\lambda = 2$, and $\lambda = 20$, respectively.

**References**

1. Sidney Coleman once said, "The career of a young theoretical physicist consists of treating the harmonic oscillator in ever-increasing levels of abstraction.". https://en.wikiquote.org/wiki/Sidney_Coleman

2. R. P. Feynman, R. B. Leighton and M. L. Sands, *The Feynman lectures on physics. Volume III, [Quantum mechanics]*, New millennium edition ed. (Basic Books, New York, 2010).

3. D. Chandler, *Introduction to modern statistical mechanics*. (Oxford University Press, New York, 1987).

4. R. Zwanzig, *Nonequilibrium Statistical Mechanics*. (Oxford University Press, New York, 2001).

5. G. Ni and S. Chen, *Advanced quantum mechanics*. (Rinton Press, Princeton, NJ, 2002).

6. R. A. Marcus, J. Elec. Chem. **438** (1-2), 251-259 (1997).

7. R. P. Feynman, F. L. Vernon and R. W. Hellwarth, J. Appl. Phys. **28** (1), 49-52 (1957).

8. B. Daino, G. Gregori and S. Wabnitz, Opt. Lett. **11** (1), 42-44 (1986).

9. R. L. Stratonovich, Zh. Eksp. Teor. Fiz. **31**, 1012 (1956).

10. J. Liu, J. Chem. Phys. **145** (20), 204105 (2016).

11. J. Liu, J. Chem. Phys. **146** (2), 024110 (2017).

12. X. He and J. Liu, J. Chem. Phys. **151** (2), 024105 (2019).

13. X. He, Z. Gong, B. Wu and J. Liu, J. Phys. Chem. Lett. **12** (10), 2496-2501 (2021).

14. X. He, B. Wu, Z. Gong and J. Liu, J. Phys. Chem. A **125** (31), 6845-6863 (2021).

15. J. Liu, X. He and B. Wu, Acc. Chem. Res. **54** (23), 4215-4228 (2021).

16. X. He, B. Wu, Y. Shang, B. Li, X. Cheng and J. Liu, Wiley Interdiscip. Rev. Comput. Mol. Sci. **12** (6), e1619 (2022).

17. Y. Shang, B. S. Thesis, Peking University, 2022.

18. B. Wu, X. He and J. Liu, J. Phys. Chem. Lett. **15**, 644-658 (2024).

19. M. Nakahara, *Geometry, Topology, and Physics*, 2 ed. (Institute of Physics

Publishing, Bristol, 2003).

20. M. F. Atiyah and J. A. Todd, Math. Proc. Cambridge Philos. Soc. **56** (4), 342-353 (1960).

21. P. A. M. Dirac, Proc. R. Soc. London, Ser. A **114** (767), 243-265 (1927).

22. H.-D. Meyer and W. H. Miller, J. Chem. Phys. **70** (7), 3214-3223 (1979).

23. S. J. Cotton and W. H. Miller, J. Chem. Phys. **150** (19), 194110 (2019).

24. J. Schwinger, in *Quantum Theory of Angular Momentum*, edited by L. C. Biedenharn and H. VanDam (Academic, New York, 1965).

25. J. J. Sakurai, *Modern Quantum Mechanics*. (Addison-Wesley, New York, 1994).

26. G. Stock and M. Thoss, Phys. Rev. Lett. **78** (4), 578-581 (1997).

27. X. Sun, H. Wang and W. H. Miller, J. Chem. Phys. **109** (17), 7064-7074 (1998).

28. S. J. Cotton and W. H. Miller, J. Chem. Phys. **145** (14), 144108 (2016).

29. W. H. Miller and S. J. Cotton, Faraday Discuss. **195**, 9-30 (2016).

30. S. J. Cotton and W. H. Miller, J. Chem. Phys. **150** (10), 104101 (2019).

31. U. Muller and G. Stock, J. Chem. Phys. **111** (1), 77-88 (1999).

32. A. A. Golosov and D. R. Reichman, J. Chem. Phys. **114** (3), 1065-1074 (2001).

33. A. A. Kananenka, C. Y. Hsieh, J. S. Cao and E. Geva, J. Phys. Chem. Lett. **9** (2), 319-326 (2018).

34. X. Gao and E. Geva, J. Phys. Chem. A **124** (52), 11006-11016 (2020).

35. Y. Xie, J. Zheng and Z. Lan, J. Chem. Phys. **149** (17), 174105 (2018).

36. E. A. Coronado, J. Xing and W. H. Miller, Chem. Phys. Lett. **349** (5-6), 521-529 (2001).

37. N. Ananth, C. Venkataraman and W. H. Miller, J. Chem. Phys. **127** (8), 084114 (2007).

38. S. J. Cotton and W. H. Miller, J. Chem. Phys. **139** (23), 234112 (2013).

39. W. H. Miller and S. J. Cotton, J. Chem. Phys. **145** (8), 081102 (2016).

40. S. J. Cotton, R. Liang and W. H. Miller, J. Chem. Phys. **147** (6), 064112 (2017).

41. R. Liang, S. J. Cotton, R. Binder, R. Hegger, I. Burghardt and W. H. Miller, J. Chem. Phys. **149** (4), 044101 (2018).

42. D. Tang, W.-H. Fang, L. Shen and G. Cui, Phys. Chem. Chem. Phys. **21** (31), 17109-17117 (2019).

43. J. Peng, Y. Xie, D. Hu and Z. Lan, J. Chem. Phys. **154** (9), 094122 (2021).

44. D. Hu, Y. Xie, J. Peng and Z. Lan, J. Chem. Theory. Comput. **17** (6), 3267-3279 (2021).

45. J. Zheng, J. Peng, Y. Xie, Y. Long, X. Ning and Z. Lan, Phys. Chem. Chem. Phys. **22** (32), 18192-18204 (2020).

46. J. Zheng, Y. Xie, S. Jiang, Y. Long, X. Ning and Z. Lan, Phys. Chem. Chem. Phys. **21** (48), 26502-26514 (2019).

47. K. Lin, J. Peng, C. Xu, F. L. Gu and Z. Lan, J. Phys. Chem. Lett. **13** (50), 11678-11688 (2022).

48. Z. Hu and X. Sun, J. Chem. Theory. Comput. **18** (10), 5819-5836 (2022).

49. Z. Hu, D. Brian and X. Sun, J. Chem. Phys. **155**, 124105 (2021).

50. Y. Liu, X. Gao, Y. Lai, E. Mulvihill and E. Geva, J. Chem. Theory. Comput. **16** (7), 4479-4488 (2020).

51. P. Linz, in *Analytical and Numerical Methods for Volterra Equations* (Society for Industrial and Applied Mathematics, Philadelphia, 1985).

52. B. Wu, X. He and J. Liu, in *Volume on Time-Dependent Density Functional Theory: Nonadiabatic Molecular Dynamics*, edited by C. Zhu (Jenny Stanford Publishing, New York, 2022).

53. B. L. van der Waerden, *Die Gruppentheoretische Methode in der Quantenmechanik*. (Springer Berlin, Heidelberg, 1932).

54. M. A. Nielsen and I. L. Chuang, *Quantum Computation and Quantum Information*. (Cambridge University Press, Cambridge, 2000).

55. A. Y. Kitaev, A. H. Shen and M. N. Vyalyi, *Classical and Quantum Computation*. (American Mathematical Society, 2002).

56. J. C. Varilly, BiBoS preprint, 345 (1989). https://kerwa.ucr.ac.cr/handle/10669/86540.

57. X. He, X. Cheng, B. Wu and J. Liu, J. Phys. Chem. Lett. **15,** 5452–5466 (2024).

https://doi.org/10.1021/acs.jpclett.4c00793.

58. F. Klein and A. Sommerfeld, *Theorie des Kreisels*. (B. G. Teubner, Leipzig, 1897).